\documentclass[11pt,3p]{scrartcl}
 
\usepackage[hidelinks]{hyperref}

\usepackage[pdftex,x11names]{xcolor}%

\usepackage{amsmath,amssymb,amsfonts}%

\usepackage{chngcntr}
 
\counterwithin{figure}{section}
\counterwithin{table}{section}

\usepackage{textcomp}%
\usepackage{manyfoot}%
\usepackage{booktabs}%
\usepackage{listings}%

\usepackage{hyperref}
\usepackage[centerdot]{mathtools}
\usepackage[T1]{fontenc}
\usepackage[utf8]{inputenc}
\usepackage{lmodern}
\usepackage[english]{babel}

\usepackage{amsthm}
\usepackage[]{mathtools}
\usepackage{dsfont}
\newcommand{\R}{\mathds{R}}
\newcommand{\Z}{\mathds{Z}}
\newcommand{\N}{\mathds{N}}
\newcommand{\X}{\mathcal{X}}
\newcommand{\Y}{\mathcal{Y}}

\newcommand{\strictly}{}

\usepackage{natbib}
\AtEndOfPackage{%
   \renewcommand\@openbib@code{%
      \advance\leftmargin\bibindent
      \itemindent -\bibindent
      \listparindent \itemindent
      \parsep \z@
      }%
   }%

\AtEndOfClass{\RequirePackage{natbib}%
\setlength{\bibhang}{\parindent}%

\def\@biblabel#1{#1.}%
\bibpunct{(}{)}{;}{a}{}{,}}

\usepackage{pifont}%
\newcommand{\cmark}{\text{\ding{51}}}%
\newcommand{\xmark}{\text{\ding{55}}}%

\DeclareMathOperator{\conv}{conv}

\DeclareMathOperator{\argmin}{argmin}

\usepackage[ruled,noend]{algorithm2e}
\SetKw{Continue}{continue}
\SetKw{Null}{null}
\SetKw{Break}{break}

\SetCommentSty{mycommfont}

\usepackage[textsize=tiny]{todonotes}
\usepackage[normalem]{ulem}
\usepackage{float}
\usepackage{subcaption}
\usepackage{authblk}
\usepackage{rotating}

\usepackage{tikz,tikz-3dplot}
\tikzset{
    cross/.pic = {
    \draw[rotate = 45] (-#1,0) -- (#1,0);
    \draw[rotate = 45] (0,-#1) -- (0, #1);
    }
}
\usetikzlibrary{shapes.misc}
\usetikzlibrary{patterns}
\usetikzlibrary{graphs,quotes}

\usepackage{pgfplots}
 \pgfplotsset{compat=1.16,
    every axis/.append style={
        axis lines=center,
        xlabel style={anchor=south west},
        ylabel style={anchor=south west},
        zlabel style={anchor=south west},
        tick align=outside,}
}
 \usepgfplotslibrary{patchplots}
\tdplotsetmaincoords{60}{120}

\usepackage[]{cleveref} 

\newtheoremstyle{theorem}
{15pt}
{5pt}
{\itshape}
{}
{\bfseries}
{}
{0.5em}
{}
\theoremstyle{theorem}

\usepackage{xpatch}

\xpatchcmd{\proof}{\itshape}{\prooflabelfont}{}{}
\newcommand{\prooflabelfont}{\bfseries}

\newtheorem{theorem}{Theorem}[section]
\newtheorem{lemma}[theorem]{Lemma}
\newtheorem{definition}[theorem]{Definition}
\newtheorem{property}[theorem]{Property}
\newtheorem{example}[theorem]{Example}

\title{ Output-sensitive Complexity of Multi-Objective Integer Network Flow Problems}

\author[1,2]{David K\"onen}
\author[1,3]{Michael Stiglmayr}
\affil[1]{%
	University of Wuppertal\\
	School of Mathematics and Natural Sciences\\
	Optimization Group\\
	Gaußstraße 20, 42103 Wuppertal, Germany\\ 
}
\affil[2]{Corresponding author, E-Mail:~\href{mailto:koenen@uni-wuppertal.de}{koenen@uni-wuppertal.de}, ORCID-ID: 0000-0003-1747-8791}

\affil[3]{E-Mail:~\href{mailto:stiglmayr@uni-wuppertal.de}{stiglmayr@uni-wuppertal.de}, ORCID-ID: 0000-0003-0926-1584 }

\date{}

\begin{document}

\maketitle

	\begin{abstract}\small
      This paper addresses the \emph{output-sensitive} complexity for linear multi-objective minimum cost integer flow problem (MMCIF), providing insights into the time complexity for enumerating all supported nondominated vectors. 
    The paper shows that there cannot  exist an output-polynomial time algorithm for the enumeration of all supported nondominated vectors that determine the vectors in an lexicographically ordered way in the outcome space unless $\mathbf{P}=\mathbf{N P}$. 
     Moreover,  novel methods for identifying supported nondominated vectors in bi-objective minimum cost integer flow problems (BMCIF) are proposed, accompanied by a numerical comparison between decision- and objective-space methods. A novel, equivalent, and more compact formulation of the minimum cost flow ILP formulation used in the $\varepsilon$-constraint scalarization approach is introduced, demonstrating enhanced efficiency in the numerical tests.
	
\end{abstract}


	
	\par\vskip\baselineskip\noindent

\newpage

\section{Introduction}\label{Intro}

The \emph{Multi-Objective Minimum Cost Integer Flow Problem} (MMCIF)  can be written as
\begin{equation}\def\arraystretch{1.4}\label{eq:MMCIF}\tag{MMCIF}
  \begin{array}{rr@{\extracolsep{0.7ex}}c@{\extracolsep{0.7ex}}lll}
    \min  &\multicolumn{1}{l}{ c(f) }\\
    \text{s.t.}  &\displaystyle \sum_{j:(i,j)\in A}f_{ij} - \sum_{j:(j,i)\in A} f_{ji} &=& b_i 	\quad  & i\in V\\
     &   l_{ij} \leq f_{ij} &\leq& u_{ij} \quad   & (i,j)\in A \\ 
    & f &\in& \Z^m_{\geqq}  
   \end{array}
\end{equation}
in a directed graph $D=(V,A)$  with $n$ nodes $V=\{v_1,\ldots,v_n\}$ and $m$ arcs $A=\{e_1,\ldots,e_m\}$. Let $l_{ij}$ and $u_{ij}$ denote the integer-valued, non-negative, finite lower and upper capacity bounds, respectively, for each arc $(i,j)\in A$. 
Here, $b_i$ is the integer-valued \emph{flow balance} of the node $i\in V$ where $b_i>0$ are supply nodes, $b_i<0$ are demand nodes, and $b_i=0$ are transshipment nodes.  

We denote a feasible solution $f=(f_{ij})_{ij\in A} \in \Z^m_{\geqq}$ as \emph{flow} and define the \emph{cost} of a flow $f$ using  the (vector valued) objective function $c\colon\, \R^m \rightarrow \R^d_{}$ and the corresponding cost matrix $C\in\R^{d\times m}$  as  
\[
  c(f)\coloneqq \bigl(c^1(f),\dots,c^d(f)\bigr)^\top  \coloneqq  \Bigl(\sum_{(i,j) \in A} c^1_{ij}\, f_{ij}, \dots, \sum_{(i,j) \in A} c^d_{ij} \,f_{ij}\Bigr)^\top =C\cdot f.
\]

 The \emph{(continuous) Multi-Objective Minimum Cost Flow Problem} (MMCF) is the LP-relaxation of MMCIF.  Note that MMCF is a totally unimodular problem, i.e., every extreme point of its feasible set is integer. From now on, we assume that $D$ is connected  (see~\citealt{Ahuja1993}) and that there is at least one feasible flow.

Then, we denote the set of feasible outcome vectors in the outcome space by $\Y \coloneqq   \{C\, f \colon f\in \X \}$, where $\X$ is the set of all feasible flows. 
We assume that the objective functions are conflicting, implying that no solution simultaneously minimizes all objectives. 
Throughout this article, we will use the \emph{Pareto concept of optimality}, which is based on the component-wise order in $\R^d$. We write $y^{1} \leqq y^{2}$ if $y_{k}^{1} \leqq y_{k}^{2}$ for $k=1, \ldots, d, y^{1} \leq y^{2}$ if $y^{1} \leqq y^{2}$ and $y^{1} \neq y^{2}$, and $y^{1}<y^{2}$ if $y_{k}^{1}<y_{k}^{2}, k=1, \ldots, d$. Furthermore, let  $\R^d_{\geqq} \coloneqq  \{x\in \R^d \colon x \geqq 0\} $ denote the non-negative orthant of $\R^d$. Its interior $\R^d_{>}$ is defined accordingly.

A feasible solution $f^{*} \in \X$ is \emph{efficient} if there does not exist any other feasible solution $f \in \X$ such that $c(f) \leq c\left(f^{*}\right)$. If $f^{*}$ is efficient, $c\left(f^{*}\right)$ is called  \emph{nondominated point}. If $f, f^{\prime} \in \X$ are such that $c(f) \leq c\left(f^{\prime}\right)$ we say that $f$ \emph{dominates} $f^{\prime}$ and $c(f)$ \emph{dominates} $c\left(f^{\prime}\right)$. Feasible solutions $f, f^{\prime} \in \X$ are \emph{equivalent} if $c(f)=$ $c\left(f^{\prime}\right)$. The set of efficient flows is denoted by $\X_E\subseteq \X$ and the set of nondominated vectors by $\mathcal{Y_N}\subseteq \Y$. Moreover, a feasible flow $f$ is called weakly efficient if there is no other flow $f^{\prime}$ such that $c({f}^{\prime}) < c(f)$.

For the weighted sum scalarization, we define the set of \emph{normalized weight vectors} as the set $\Lambda_d=\{\lambda \in \R^{d}_{>} \colon \|\lambda\|_{1} = 1 \}$ or $\Lambda_d^0=\{\lambda \in \R^{d}_{\geq} \colon \|\lambda\|_{1} = 1 \}$ if weights equal to zero are included. The \emph{weighted sum scalarization} with $\lambda \in \Lambda_d$ or $\lambda \in \Lambda_d^0$ is defined as the parametric program $ P_{\lambda}\coloneqq  \min \{\lambda^\top  Cf \colon f \in \X\} $.

If $\lambda\in \Lambda_{d}^0$, every optimal solution of $P_{\lambda}$ is a weakly efficient solution. Moreover, every optimal solution of $P_{\lambda}$ is efficient if \(\lambda\in\Lambda_{d}\)~\citep{ehrgott2005multicriteria}.

Then, we can distinguish between several classes of efficient solutions. 
\begin{enumerate}
\item \emph{Weakly supported efficient solutions} are efficient solutions that are optimal solutions of $P_{\lambda}$ for  $\lambda\in \Lambda_{d}^0$, i.e., an optimal solution to a weighted sum single-objective scalarization with weights strictly or equal to zero. Their images in the outcome space are weakly \emph{supported nondominated points}. All weakly supported nondominated vectors are located on the boundary of the \emph{upper image} $\mathcal{P}\coloneqq  \operatorname{conv}(\Y)+ \R_\ge^d.$

\item An efficient solution is called \emph{supported efficient solution} if it is an optimal solution of a weighted sum scalarization $P_{\lambda}$ with $\lambda\in \Lambda_{d}$, i.e., an optimal solution to a weighted sum scalarization where the weights are strictly positive. Its image is called \emph{supported nondominated vector}; we use the notation $\X_{SE}$ and $\Y_{SN}$. A supported nondominated vector is located on the \emph{nondominated frontier} defined as the set $\{y \in \conv(\Y_N) \colon \conv(\Y_N) \cap (y -\R^d_{\geq}) = \{y\}\}$, i.e., located on the union of the \emph{maximal nondominated faces}. 
A \emph{nondominated face} $F\subset \Y$ is a face of the feasible outcome sets $\Y$ such that all points on the face $F$ are nondominated.
A face $F$ is called \emph{maximally nondominated} if there is no other nondominated face $G$ such that $F$ is a proper subset of $G$.  Let $F_\X$ be the preimage of a maximally nondominated face $F_\Y$ of polytope $\Y$. In slight abuse of notation, we refer to $F_\X$ as \emph{maximally efficient face}, even though $F_\X$ does not necessarily correspond to a face of the feasible set in decision space.

\item \emph{Extreme supported solutions}  are those solutions whose image lies on the vertex set of the upper image. Their image is called an \emph{extreme supported nondominated vector}. We use the notation $\Y_{EN}$ for the set of extreme nondominated points.

\item \emph{Unsupported efficient solutions} are efficient solutions that are not optimal solutions of  $P_{\lambda}$ for any  $\lambda\in \Lambda_{d}^0$. \emph{Unsupported nondominated vectors} lie in the interior of the upper image. 
\end{enumerate}

 ~\Cref{Image:DiffSolutions} illustrates supported extreme, supported, and unsupported nondominated points as well as the upper image in the bi-objective case. The distinction between supported and weakly supported solutions is relatively new,
 and a comprehensive introduction can be found in~\cite{koenen2025supportednessmultiobjectivecombinatorialoptimization}. This distinction provides a consistent definition of supportedness in MOCO
problems.

\begin{figure}
    \centering
    \begin{tikzpicture}[]
        \tdplotsetmaincoords{0}{0}
        \tikzstyle{vertex}=[circle,fill=black,draw=black,minimum size=4pt,inner sep=0]
        \tikzstyle{vertex2}=[circle,draw=DeepPink4,minimum size=4pt,inner sep = 0]
        \tikzstyle{vertex3}=[rectangle,fill=DodgerBlue4,draw=DodgerBlue4,minimum size=4pt,inner sep = 0]
        \tikzstyle{N_point}=[draw, cross out,scale=.5,thick]

        \begin{scope}[tdplot_main_coords, scale=.4]
          
        \draw [draw= white, fill= black!10  ,fill opacity=1 ]  
        (2,12.25) -- (2,9) -- (3,6)  -- (8,3) -- (12.5,3) -- (12.5,12.25)-- cycle;
	
        \draw[] (-1,0) -- (10,0);
        \draw[] (0,-1) -- (0,10);
    	\draw[->] (10,0) node[anchor=north]{} -- (13,0) node[anchor=north]{\small$c^1(f)$};
    	\draw[->] (0,10) -- (0,13) node[anchor=north east]{\small$c^2(f)$};

        \draw[thick] (2,9) -- (3,6) -- (8,3);
        \draw[dotted,thick] (2,9) -- (2,12.25);
        \draw[dotted,thick] (8,3) -- (12.5,3);
         
        \node[] at ( 8.5 ,11 ){\small $ \conv(\Y+\R^2_{\geqq}) $};

        \node[vertex3, label={[label distance=-5pt]above right:{}}] at (2,9) {};
        \node[vertex3, label={[label distance=-5pt]above right:{}}] at (3,6) {};
        \node[vertex, label={[label distance=-5pt]above right:{}}] at (6,5) {};
        \node[vertex3, label={[label distance=-5pt]above right:{}}] at (8,3) {};
        \node[vertex2, label={[label distance=-5pt]above right:{}}] at (4,5.4) {};
        \node[vertex2, label={[label distance=-5pt]above right:{}}] at (7,3.6) {};
        \node[vertex2, label={[label distance=-5pt]above right:{}}] at (2.5,7.5) {};
        \node[N_point, label={[label distance=-5pt]above right:{}}] at (3,9) {};
        \node[N_point, label={[label distance=-5pt]above right:{}}] at (7,7) {};
  
        \draw (15,3.25) -- (15,9) -- (29,9) -- (29,3.25) -- cycle;
        \node[vertex3] at (16,8) {};
         \draw (16.5,8) node [anchor= west][inner sep=0.75pt]   [align=left] {\scriptsize extreme supported non-dominated};
         \node[vertex2] at (16,6.75) {};
         \draw (16.5,6.75) node [anchor= west][inner sep=0.75pt]   [align=left] {\scriptsize supported non-dominated};
         \node[vertex] at (16,5.5) {};
         \draw (16.5,5.5) node [anchor= west][inner sep=0.75pt]   [align=left] {\scriptsize unsupported non-dominated};
         \node[N_point] at (16,4.25) {};
         \draw (16.5,4.25) node [anchor= west][inner sep=0.75pt]   [align=left] {\scriptsize dominated};
        \end{scope}
        
    \end{tikzpicture}
\caption{Illustration of the upper image $\mathcal{Y}^{\geqq}=\conv(\mathcal{Y})+ \R^2_{\geqq}$ and the different solution types.}\label{Image:DiffSolutions}\end{figure}

Due to the total unimodularity of MMCF, each extreme supported nondominated point of MMCF has an integer preimage since $u$ and $b$ are integral (see~\citealt{Ahuja1993}). In other words, the sets of extreme supported nondominated points of MMCF and MMCIF, and thus the respective upper images, coincide.
In the remainder of this paper, only integer flows are considered, and from now on, flow always refers to an integer flow.  

The minimum cost flow problem is a fundamental, well-studied problem in combinatorial optimization~\citep{Ahuja1993,bertsekas98network} and various polynomial algorithms exist for its single-objective version. However, real-world problems often involve multiple conflicting objectives. Consider, for example, a transportation problem aiming to minimize costs, environmental impact, and  time while maximizing safety.  These multi-objective minimum cost flow problems are much more complex than the single-objective case.
 While MMCF problems contain only supported efficient solutions,
the efficient solution set of MMCIF also contains unsupported efficient solutions. Thereby, the unsupported efficient solutions typically outnumber
the supported ones as observed in~\cite{Visee1998} for general MOCO problems. The computation of unsupported non-dominated points requires different scalarization techniques  than the weighted sum scalarization and is often computationally more expensive. Furthermore, the supported nondominated vectors can already serve as high-quality
representations~\citep{Serpil2024}, and can be used as a foundation for two-phase methods to generate the complete nondominated point set.
MMCIF had been reviewed in~\cite{Hamacher2007}, where the authors comment on the lack of computationally efficient algorithms.

There are several algorithms to determine efficient solutions for bi-objective minimum cost integer flow problems (BMCIF), e.g.,~\citet{Eusebio09,eusebio14,raith09,sedeno-noda01,sedona00,sedona03}. Raith and Sede\~no-Noda introduced an enhanced parametric approach to determine all extreme efficient solutions for BMCIF~\citep{raith17}.
 However, there are only quite a few specific methods designed to determine all (or subsets) of the nondominated vectors in the outcome space (nor the corresponding efficient solutions in the decision space) for MMCIF~\citep{EUSEBIO200968,fonesca10,sun11}.

 \cite{konen2023outputpolynomial} present an \emph{output-polynomial} time algorithm for determining all efficient supported flows for MMCIF.  
Unfortunately, this approach is insufficient for computing all nondominated supported vectors in output-polynomial time, as a single vector may correspond to an exponential number of flows.

Analyzing the output-sensitive complexity of specific problems has gained importance in recent years. Several combinatorial problems have been studied, e.g., multi-objective shortest path~\citep{Boekler2017}, multi-objective spanning tree problems~\citep{Boekler2017,OKAMOTO201148}, mining frequent closed attribute trees~\citep{10.1007/11536314_1}, general multi-objective combinatorial optimization problems, and multi-objective linear programs~\citep{Boekler2015}.

The purpose of this paper is to give insights into the
time complexity for the enumeration of all supported nondominated vectors for MMCIF. The
paper shows that there cannot  exist an output-polynomial time algorithm for the
enumeration of all supported nondominated vectors that determine the vectors in
an ordered way in the outcome space unless $\mathbf{P}=\mathbf{NP}$. 
 The paper shows that the next best distinct cost flow can be determined in $\mathcal{O}(n^3)$ given an initial optimal flow.  This result derives an improved algorithm to determine all supported nondominated vectors for BMCIF to the algorithm presented in~\cite{konen2023outputpolynomial} if the number of branches needed is significantly smaller than the number of supported efficient solutions.  The paper presents an example in which the adjusted algorithm saves an exponential amount of considered flows to get all supported nondominated vectors. 

While the algorithm in~\cite{konen2023outputpolynomial} is a \emph{decision-space} method, the paper presents also \emph{objective-space} methods to determine all supported nondominated vectors, based on the $\varepsilon$-constraint scalarization. In addition, the paper gives an equivalent, more compact formulation for the ILP used in the $\varepsilon$-constraint method, which can be solved more efficiently.  The $\varepsilon$-constraint scalarization could also be used to get all nondominated vectors, even in higher dimensions. The adjusted algorithm can be seen as a combination of a decision- and objective-space method. 

 The remainder of the paper is structured as follows.~\Cref{sec:output} introduces the theory of output-sensitive time complexity and summarizes the existing results from the literature on the existence of output-polynomial time algorithms for the MMCIF w.r.t.\ different solution concepts. \Cref{output-MMCIF} provides insight about the output-sensitive complexity for MMCIF. In~\Cref{sec:algo}, we derive an adjusted algorithm to determine all supported nondominated vectors. In addition, a method is presented to use objective-space methods like the $\varepsilon$-constraint scalarization to determine all supported nondominated vectors, and a more compact formulation for the ILP used in the $\varepsilon$-constraint scalarization is presented. Numerical results by these algorithms on different instances are reported in~\Cref{chapt:Numerical}.~\Cref{chapt:concl} summarizes the paper's findings and suggests potential avenues for future research.

\section{An Introduction to Output-sensitive Complexity}
\label{sec:output}

This section formally introduces the theory of \emph{output-sensitive} complexity of enumeration problems and summarizes the 
existing results from the literature on the existence of output-polynomial time algorithms for the MMCIF w.r.t.\ different solution concepts. For a comprehensive introduction, see e.g.~\cite{JOHNSON1988119}.

\begin{definition}[\citealt{Boekler2017}]
     An enumeration problem is a pair $(I, C)$ such that 
\begin{itemize}
    \item[1.] $I \subseteq \Sigma^*$ is the set of \emph{instances} for some fixed alphabet $\Sigma$,
    \item[2.] $C: I \rightarrow 2^{\Sigma^*}$ maps each instance $x \in I$ to its \emph{configurations} $C(x)$, and
    \item[3.] the encoding length $|s|$ for $s$ in $C(x)$ for $x$ in $I$ is in $\operatorname{poly}(|x|)$,
    \end{itemize}
    where $\Sigma^*$ can be interpreted as the set of all finite strings over $\{0,1\}$. 
\end{definition}
We assume that $I$ is decidable in polynomial time and that $C$ is computable.

\begin{definition}[\citealt{Boekler2017}]
   An \emph{enumeration algorithm} for an enumeration problem $E=(I, C)$ is a \emph{random access machine} that
   \begin{itemize}
    \item on input $x \in I$ outputs each $c \in C(x)$ exactly once, and
    \item on every input terminates after a finite number of steps.
   \end{itemize}
\end{definition}

\begin{definition}[\citealt{Boekler2017}]
    An enumeration algorithm for an enumeration problem $ E=(I, C)$ is said to run in \emph{output-polynomial time} \emph{(is output-sensitive)} if its running time is in $\operatorname{poly}(|x|,|C(x)|)$ for $x$ in $I$.
\end{definition}

A finished decision problem  $E^{\mathrm{Fin}}$ for an enumeration problem $E=(I, C)$ is defined as the problem: Given an instance $x \in I$ of the enumeration problem and a subset $M \subseteq C(x)$ of the configuration set, the goal is to decide if $M=C(x)$, i.e.,  we want to decide if we already have found all configurations.  If the enumeration problem $E$ can be solved in output-polynomial time then $E^{\mathrm{Fin}}\in \mathbf{P}$~\citep{Lawler1980}.

Thus, MMCIF, as any discrete multi-objective optimization problem, can be considered an enumeration problem, i.e., enumerating all nondominated vectors (or w.r.t.\ different solution concepts subsets of the nondominated vectors or the efficient solutions). Let $C$ be the configuration set of all supported efficient solutions and $C^*$ the configuration set of all nondominated vectors. An output-polynomial time algorithm $E=(I,C)$, which determines all supported efficient solutions exactly once, would also yield all nondominated vectors. However, it might not be output-polynomial w.r.t.\ this task, as there might be an exponential number of efficient solutions mapping to a small number of nondominated points.

Thus, the algorithm for $E^*=(I,C^*)$  may output the elements (i.e., the nondominated vectors) more than once or even exponentially many times. 
Consider, for example, a directed graph with $\{1,\dots,n\}$ transshipment nodes, a node $s$ and $t$ with flow balance $n$ and $-n$, respectively. The graph contains the arcs $(s,i)$ and $(i,t)$ for all  $i \in\{1,\dots,n\}$ with upper capacity $n$. The cost of all arcs is equal to \((1,1)^\top\in\R^2\). Then, we have $\binom{2n-1}{n}$ supported efficient solutions, but all map to the same extreme nondominated point. 

\Cref{tab1} summarizes the existing results from literature on the existence of output-polynomial time algorithms for the MMCIF w.r.t.\ different solution concepts. In the table, a check mark indicates existence, a cross indicates that the existence of such an algorithm can be ruled out, and the question mark indicates that this problem remains an open question.

\begin{table}[htb]\footnotesize
\caption{Existing results on the existence of output-polynomial time algorithms for the MMCIF w.r.t.\ different solution concepts.}\label{tab1}
\begin{center}
\begin{tabular}{p{7em}|cccc}
\toprule
& \parbox{5em}{extreme supported}\rule[-1em]{0pt}{2.5em} & supported & \parbox{5em}{weakly supported} & \text{\quad all\quad} \\\midrule
\raisebox{0.25ex}{\parbox{5em}{nondominated\linebreak vectors}} & \cmark\footnotemark[1] &   \textbf{?} & \xmark\footnotemark[2]  &  \xmark\footnotemark[3] \\\midrule
\raisebox{0.25ex}{\parbox{5em}{efficient\linebreak solutions}} & \cmark & \cmark\footnotemark[2] & \xmark\footnotemark[2]  & \xmark \\
\bottomrule
\multicolumn{4}{p{8cm}}{\scriptsize
\textsuperscript{1}\; \cite{Boekler2015,Ehrgott2012}\rule{0pt}{4ex}\newline
\textsuperscript{2}\; \cite{konen2023outputpolynomial}\newline
\textsuperscript{3}\; \cite{Boekler2017}
}
\end{tabular}
\end{center}
\end{table}

Note that the negative results for all nondominated vectors also hold for the bi-objective case and the weakly supported nondominated vectors for the three-objective case. In the bi-objective case, the set of all weakly supported nondominated vectors equals the set of all supported nondominated vectors~\citep{konen2023outputpolynomial}. While an output-polynomial time algorithm exists to determine all supported efficient solutions, it  remains an open question for the set of all supported nominated vectors.

\section{Output-Sensitive Complexity for MMCIF}
\label{output-MMCIF}
This chapter provides new insights into the time complexity for enumerating all supported nondominated vectors. We show that  no output-polynomial time algorithm for enumerating all supported nondominated vectors that finds the vectors in an lexicographically ordered way in the outcome space exists unless $\mathbf{P} = \mathbf{NP}$. In order to prove this result, we first need the concepts of \emph{optimal tree solution} and  \emph{induced cycles}.

\subsection{Preliminaries}
\begin{definition}
	We call $f$ and an associated \emph{tree structure} $(T,L,U)$ a \emph{tree solution} if it consists of  a spanning tree $T$ of $D$ and a disjoint edge set $A\setminus T= L \, \cup \, U$ with $U\cap L=\varnothing$ such that the flow $f$ satisfies
	\begin{alignat*}{2}
        &f_{ij} = l_{ij}  &\quad&  \text { for all } (i,j) \in L, \\ 
        &f_{ij} =u_{ij} & & \text { for all } (i,j) \in U. 
    \end{alignat*}
\end{definition}

If a feasible (optimal) flow exists, then there also exists a feasible (optimal) tree solution~\citep{Cook1998}, respectively. 
 The network simplex algorithm~\citep{Dantzig1951} always determines an optimal tree solution~\citep{Ahuja1993}. Note that the network simplex algorithm does not run in polynomial time. However, using the \emph{enhanced capacity scaling algorithm}~\citep{Orlin1993} solves the problem in time $O\bigl((m \log n)(m + n\log n)\bigr)$, i.e., in strongly polynomial time would yield an optimal solution that can be transformed into a tree solution in polynomial time.  
 W.l.o.g.\ we can assume that $l_{ij} = 0$ for all $(i,j)\in A$ (see~\citealt{Ahuja1993}).
 
 We associate a real number $y_i$ with each node $i\in V$. We refer to $y_i$ as the node potential of node $i$. For a given set of node potentials $y$
we define the \emph{reduced cost} $\overline{c}_{ij}$ of a given arc $(i,j)$ as $\overline{c}_{ij}\coloneqq c_{ij} + y_i - y_j$.
Any node potential $y$ that satisfies the well-known \emph{complementary slackness optimality conditions}  (see, e.g.~\citealt{Ahuja1993}) is \emph{optimal}, and we have the following property of the reduced costs in the \emph{residual graph \(D_f\)}.
\begin{property}\label{prop:complSlack}
	Let $f$ be an optimal flow and $y$ an optimal node potential. Then
	$\overline{c}_{ij}(f)\ge0$  for all $(i,j)\in A_f$.
\end{property} 

Hereby, $D_f=(V,A_f\coloneqq  A^{+}  \cup  A^{-})$ is defined as the \emph{residual graph} with respect to a feasible flow $f$, where 
$
A^+ \coloneqq \{(i,j) \colon (i,j)\in A,\, f_{ij} < u_{ij}\}\ \text{ and }  
A^- \coloneqq \{ (j,i) \colon (i,j) \in A,\, f_{ij}>l_{ij}\}.
$
In $D_f$, the \emph{residual capacities} and \emph{residual costs} are defined by  $ u_{ij}(f) \coloneqq  u_{ij}-f_{ij} > 0$ and $c_{ij}(f)\coloneqq c_{ij}$, respectively, if $(i,j)\in A^+$ and $u_{ij}(f)\coloneqq f_{ji}-l_{ji} > 0$ and $c_{ij}(f)\coloneqq -c_{ji}$, respectively, if $(i,j)\in A^-$. 

For an optimal flow $f$, we can determine an optimal node potential $y$ by computing shortest path distances in $D_f$  from an arbitrary but fixed root node $r$ to all other nodes $i\in V$. 
Notice that we assume that all nodes $i$ are reachable from $r$ in $D_f$. If this is not the case, artificial arcs with sufficiently high costs are added to $D_f$. In the following, $y$ is considered as an optimal node potential (see~\citealt{Ahuja1993}).

\begin{property}[\citealt{Ahuja1993}]
	A tree solution $f$ with an associated tree structure $(T,L,U)$ is optimal if 
	\begin{alignat*}{3}
		&(i) &\quad & \overline{c}_{ij} = 0 &\; & \text{ for all } (i,j)\in T, \\
		&(ii) & & \overline{c}_{ij} \ge 0 & & \text{ for all } (i,j)\in L, \\
		&(iii) & &  \overline{c}_{ij} \le 0 & & \text{ for all } (i,j)\in U.
	\end{alignat*}
\end{property}

These three conditions are equivalent to $\overline{c}_{ij}(f) \geq 0$ for all $(i,j) \in D_f$.

\begin{figure}[htb]
	
    \centering\footnotesize
    \begin{minipage}[t]{0.45\textwidth}
\centering
  	\tikzstyle{vertex}=[circle,fill=white,draw=black,minimum size=20pt,inner sep=0]
  	\tikzstyle{edge} = [draw,thick,->]
  	\tikzstyle{weight} = []
  	\tikzstyle{selected vertex} = [vertex, fill=red!24]
  	\tikzstyle{selected edge} = [draw,line width=5pt,-,green!50]
  	\usetikzlibrary{arrows,automata}
  	\pgfdeclarelayer{background}
  	\pgfsetlayers{background,main}
  	\begin{tikzpicture}[->,shorten >=1pt,auto,node distance=2.8cm,
  	semithick]
  	\foreach \pos/\name in {{(0,0)/1}, {(1.5,1.5)/2}, {(1.5,-1.5)/3},
  		{(3.5,1.5)/4}, {(5,0)/5}}
  	\node[vertex] (\name) at \pos {$\name$};
  	
  	\path (1) edge[->,thick]   	node[sloped] {(10,\color{DeepPink4}3\color{black},\color{DodgerBlue4}5)} (2)
  	(1) edge[->,thick]   	node[sloped,below] {(5,\color{DeepPink4}8\color{black},\color{DodgerBlue4}1)} (3)
  	(2) edge[->,thick]   	node[sloped] {(4,\color{DeepPink4}5\color{black},\color{DodgerBlue4}5)} (3)
  	(2) edge[->,thick]   	node[sloped] {(7,\color{DeepPink4}3\color{black},\color{DodgerBlue4}9)} (4)
  	(3) edge[->,thick]   	node[sloped,below] {(8,\color{DeepPink4}2\color{black},\color{DodgerBlue4}7)} (4)
  	(3) edge[->,thick]   	node[sloped,below] {(6,\color{DeepPink4}10\color{black},\color{DodgerBlue4}2)} (5)
  	(4) edge[->,thick]   	node[sloped] {(8,\color{DeepPink4}1\color{black},\color{DodgerBlue4}4)} (5)
  	; 
  	
  	\node at (0.8,0) {$-10$};
  	\node at (4.2,0) {$10$};
  	
  	\node at (2.5,-2.5) {$(u_{ij}$,\color{red}$c^1_{ij}$\color{black},\color{DodgerBlue4}$c^2_{ij}$\color{black}$)$};
\end{tikzpicture}
    \end{minipage}
     \begin{minipage}[t]{0.45\textwidth}
     \centering
		\tikzstyle{vertex}=[circle,fill=white,draw=black,minimum size=20pt,inner sep=0]
				\tikzstyle{edge} = [draw,thick,->]
				\tikzstyle{weight} = []
				\tikzstyle{selected vertex} = [vertex, fill=red!24]
				\tikzstyle{selected edge} = [draw,line width=5pt,-,yellow!20]
				\usetikzlibrary{arrows,automata}
				\pgfdeclarelayer{background}
				\pgfsetlayers{background,main}
				\begin{tikzpicture}[->,shorten >=1pt,auto,node distance=2.8cm,
				semithick]
				\foreach \pos/\name in {{(0,0)/1}, {(1.5,1.5)/2}, {(1.5,-1.5)/3},
					{(3.5,1.5)/4}, {(5,0)/5}}
				\node[vertex] (\name) at \pos {$\name$};
				
				\path (1) edge[->,thick]   	node[sloped] {(0,10,\color{DodgerBlue4} 7\color{black})} (2)
			(1) edge[->,thick]   	node[sloped,below] {(0,5,\color{DodgerBlue4} 3\color{black})} (3)
			(2) edge[->,thick,dashed]   	node[sloped] {(0,4,\color{DodgerBlue4} 0\color{black})} (3)
			(2) edge[->,thick,dashed]   	node[sloped] {(0,7,\color{DodgerBlue4} 7\color{black})} (4)
			(3) edge[->,thick]   	node[sloped,below] {(0,8,\color{DodgerBlue4} 1\color{black})} (4)
			(3) edge[->,thick]   	node[sloped,below] {(0,6,\color{DodgerBlue4}2\color{black})} (5)
			(4) edge[->,thick,dashed]   	node[sloped] {(0,8,\color{DodgerBlue4}8\color{black})} (5)
			;

				\draw[<-,thick] ([shift=(3:1mm)]4,0.2) arc (0:310:4mm);

				\begin{pgfonlayer}{background}
				\foreach \source / \dest in {3/4,3/5,4/5}
				\path[selected edge] (\source.center) -- (\dest.center);
				\end{pgfonlayer}

  	\node at (2.5,-2.5) {$(l_{ij}$,$u_{ij}$\color{black},\color{DodgerBlue4}$f_{ij}$\color{black}$)$};
				
				\end{tikzpicture} 
    \end{minipage}
    \caption{ An example of a BMCIF, that contains $10$ nondomianted vectors and $83$ non-efficient flows. Right: an optimal tree solution for $c^1$ and a example of a unique induced cycle  $C_{4,5}$ induced by the arc $(4,5)$ (non-tree arcs are dashed). Here arc $(4,5)\in U$ and therefore $C^+_{4,5}=\{(4,5),(3,4)\}$  and $C^-_{4,5}=\{(3,5)\}$. The cost $c(C_{4,5})= (7,-9)^{\top}.$}
    \label{fig:uniqueCycleCij2}
\end{figure}
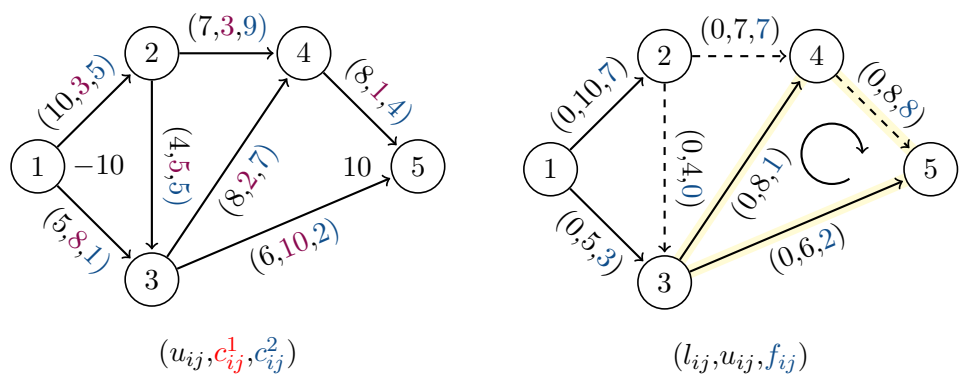

\begin{definition}[\citealt{Cook1998}]\label{def:Ca}
	Let $(i,j) \notin T$ be a non-tree arc.
	\begin{itemize}
		\item[(i)]	There exists a \emph{unique} cycle $C(T, L, U, (i,j))$  \emph{induced by }$(i,j)$, that is formed by $(i,j)$ together with the path $P^T_{ji}$ from $j$ to $i$ in $T$. This cycle is referred to as $C_{ij}$.
		\item [(ii)]  The arc $(i,j)$ defines the orientation of the cycle $C_{ij}$. If $(i,j)\in L$, then the orientation of the cycle is in the same direction as $(i,j)$. If otherwise $(i,j)\in U$, then the cycle's orientation is in the opposite direction. We define the set of all arcs $(u,v)$ in $C_{ij}$ that is in the same direction as the orientation of the cycle with $C^+_{ij}$ and the set of all arcs $(u,v)$ that are opposite directed with $C^-_{ij}$.
	\end{itemize}
	See~\Cref{fig:uniqueCycleCij2} for an illustration. 
\end{definition}

\begin{property}[\citealt{Cook1998}]\label{prop:CycleCost=ReducedCost}
	Given a tree structure $(T, L, U)$, the unique cycle $C_{ij}$ induced by an arc $(i,j) \not\in T$ satisfies
	\begin{itemize}
		\item For all arcs $(u,v)\in C_{ij}$ it holds that $(u,v) \in T\cup (i,j)$.
		\item The cost of $C_{ij}$ is given by $c(C_{ij})=\overline{c}_{ij}$ if $(i,j)\in L$ and $c(C_{ij})=-\overline{c}_{i,j}$ if $(i,j)\in U$.
	\end{itemize}
\end{property}

\begin{theorem}[\citealt{KONEN2022333}] \label{AnyCycleFrominducedCycles}
	Each undirected cycle $ C_D$ in $D$ can be represented by a composition  of the unique induced cycles $$ C_{ij} \text{ for all } (i,j)\in C_D \setminus T. $$ 
\end{theorem}

Any cycle $C\in D_f$ yields an undirected cycle $C_D$, which can be represented by a composition of the unique  cycles  $ C_{ij} \text{ for all } (i,j)\in C_D \setminus T $.
We can extend this result by showing that the incidence vector and the cost are closed under this composition.

\begin{lemma}[\citealt{KONEN2022333}]\label{lem:DirectedComposition}
	For any cycle $C\in D_f$, let $C_D$ be the undirected equivalent cycle in $D$. Then, it holds that:
	$$ \chi(C) = \sum_{a\in C_D\setminus T} \chi(C_a) \qquad\text{ and } \qquad c(C,f) = \sum_{a\in C_D\setminus T} c(C_a),$$
	where $\chi(C_a)\in \{-1,0,1\}^A$ is defined by 
	\begin{align*}
		\chi_{ij}(C_{a}) \coloneqq  \begin{cases}
			1, &\text{if $(i,j)\in C^+_{a},$ }\\
			-1, &\text{if $(i,j)\in C^-_{a},$}\\
			0, &\text{otherwise}\\
		\end{cases} && \text{for all $(i,j) \in A$.}
	\end{align*}
\end{lemma}

\begin{theorem}[\citealt{KONEN2022333}]\label{theo:optimalFCompositionOfCa}
	Let $f$ be an optimal tree solution with an associated tree structure $(T,L,U)$ and let $f^*$ be another integer flow. Then the flow $f^*$ can be written as 
	$$ f^* = f + \sum_{a \notin T} \lambda_a\, \chi(C_a)$$
	for some $\lambda \in \mathbb{Z}$ and it holds that 
        $$c(f^*)= c(f) + \sum_{a \notin T} \lambda_a\, c(C_a) $$
\end{theorem}

\subsection{Output-Sensitive Complexity for BMCIF}

In order to investigate the time complexity of enumerating all
supported nondominated vectors for BMCIF, we first define new problems in single-objective MCIF. 

\begin{definition}
    Given a single-objective \emph{Minimum Cost Integer Flow Problem} (MCIF) and an integer $k\in \mathbb{Z}$. Then the \emph{exact flow problem (EF)} asks whether there exists a flow $f$ with cost $c(f) =k$.
\end{definition}

We will prove that this decision
problem is $\mathbf{NP}$-complete by reducing it to the well-known $\mathbf{NP}$-complete \emph{subset sum} problem. 

\begin{definition}
     Given a set $N=\{1, \ldots, n\}$ of $n$ items with positive integer weights $w_1, \ldots, w_n$ and a real value $T$, the \emph{subset sum problem} is to find a subset of $N$ such that the corresponding total weight is exactly equal  to $T$. The formal definition  is given by 
    \begin{align*}
      \sum_{j=1}^n w_j \,x_j &= T\\
      x_j &\in\{0,1\} \quad \forall j\in\{1, \ldots, n\}
    \end{align*}
\end{definition}

\begin{theorem}\label{thm:cvf-np}
The exact flow problem is $\mathbf{NP}$-complete.
\end{theorem}
\begin{proof}

Take an instance of the subset sum problem.
Create the following instance of an exact flow problem. Create a node $i$
for each $i \in N$ and an additional node $n+1$. Create two arcs $\left(i, i+1\right)\; \forall i\in\{1,\ldots, n\}$, one with cost equal to zero and one with cost equal to $w_i$. All arcs get lower and upper capacities equal to zero and one, respectively.  We define the nodes $1=s$ and  $n+1=t$. We set $b_s=-1,b_t=1$  and $b_i=0$ for all other nodes.  A construction of this instance is illustrated in~\Cref{fig:proof_instance}. Note that artificial nodes and arcs could be added to prevent multi-arcs.

If we can decide whether a flow $f\in \X $ with $ c(f)  = T $  exists in polynomial time, we would also be able to solve the subset sum problem in polynomial time. Assume the subset sum problem is solvable, i.e., $\sum_{j=1}^n w_j x_j = T$. For each $x_i=1$ we take the arc $(i,i+1)$ with cost $c_{i,i+1}=w_i$. For each $x_i=0$, we take the arcs $(i,i+1)$  with a cost equal to zero. This is a feasible $s$-$t$  path (feasible flow) with a cost equal to $T$. On the other side, considering a path $P$ (flow)  with cost equal to $T$, we set $x_i=1$ if $(i,i+1)\in P$ that have cost equal $w_i$ and $x_i=0$ if not. Then $x$ solves the subset problem with the same argumentation as above.

\end{proof}

\tikzstyle{vertex}=[circle,fill=black,draw=black,minimum size=8pt,inner sep=0]
\tikzstyle{svertex}=[circle,fill=black,draw=black,minimum size=5pt,inner sep=0]%
\tikzstyle{wvertex}=[circle,fill=white,draw=white,minimum size=8pt,inner sep=0]
\begin{figure}[htb]
	\centering
	\footnotesize
	\begin{tikzpicture}[ scale=0.9,->,shorten >=1pt,auto,node distance=2.6cm,%
		semithick]%
		\foreach \stil/\pos/\name/\label in {{vertex/(-2,0)/a/s=1}, {vertex/(2,0)/c/2},%
		   {vertex/(6,0)/e/3}, {wvertex/(8,0)/f/}, {vertex/(10,0)/g/t}}%
		\node[\stil] (\name) [label=$\label$] at \pos {};%

		\path (a) edge[->,thick]  node[sloped] {\small$0$}  	 (c)%
		(c) edge[-latex,thick]   node[sloped] {\small$0$}  (e)%
            (e) edge[-,thick]   	 (f)%
            (f) edge[-latex,thick]   	 (g)%
		(a) edge[-latex,thick,bend right]  node[sloped,below] {\small$w_1$} (c)
		(c) edge[-latex,thick,bend right]  node[sloped,below] {\small$w_2$} (e)
            (e) edge[-latex,thick,bend right]   (g)
            ;

            \node[wvertex] (h) at (8,-0.6) {$\ldots$}; 
            \node[wvertex] (h1) at (8,0) {$\ldots$};  

	\end{tikzpicture}%
	 \caption{The instance of the exact flow problem corresponding to a given subset sum instance.}
    \label{fig:proof_instance}
\end{figure}
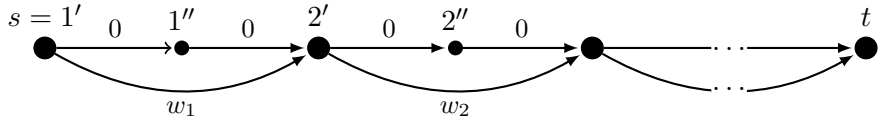

 \Cref{thm:cvf-np} implies that it is also $\mathbf{NP}$-complete to determine if a flow $f$  exists with $k_1< c(f) < k_2$ with $k_1, k_2 \in \mathbb{Z}$ .

In addition, we can prove the following statement. 

\begin{theorem}
    The problems to determine the $k$-th best flow or the $k$-th smallest distinct cost of a flow are $\mathbf{NP}$-hard.
\end{theorem}
\begin{proof}
    Let the cost of finding the $k$-th best or $k$-th smallest distinct cost of a flow be $T(n)$. Consider the instance of the exact flow problem for a given subset sum problem as in the proof above. Deciding whether a flow $f$ exists with $c(f)=T$ would solve the subset sum problem. At most, $2^n$ different flows (item $i$ could be selected or not). Thus, a binary search for a given flow value is $\mathcal{O}(n)$. It can be concluded that the complexity of the subset sum problem, known to be $\mathbf{NP}$-complete, is $\mathcal{O}(nT(n))$. Therefore, the problem of determining the $k$-th best or $k$-th smallest distinct cost of a flow is $\mathbf{NP}$-hard. 
\end{proof}

Next, we will prove that we can determine all distinct costs of the flows for a single-objective minimum cost flow problem in output-polynomial time if we can determine all supported nondominated vectors for a bi-objective minimum cost flow problem in output-polynomial time and vice versa.

\begin{definition}
Given a single-objective MCIF, the \emph{all distinct cost value flow problem (ACVF)}
determines a minimal set of flows that includes all existing different cost values of all feasible flows. In other words, it identifies $k$ different flows $ f_1, \ldots, f_k$  such that $$ c\left(f_1\right)<\ldots<c\left(f_k\right)$$ and there exists no flow $ f_p $ such that 
\(c(f_p)\notin\{c(f_1),\ldots,c(f_k)\}\).
\end{definition}

\begin{theorem}
    If we can solve the ACVF in output-polynomial time, then we can determine all supported nondominated vectors of a BMCIF in output-polynomial time.
\end{theorem}

\begin{proof}
Each supported nondominated vector lies on a maximal nondominated facet of the upper image since its preimage is an optimal solution to a weighted sum scalarization $P_{\lambda}$ for a $\lambda \in \Lambda_d$ (i.e., weights are strictly positive). Note that in the bi-objective case, every maximally nondominated face $F_\Y$ of $\conv(\Y)$ is a line segment connecting two adjacent extreme supported points if there is more than one nondominated point (\(|\Y_N|>1\)). A maximally nondominated face can only have dimension zero if there is only one extreme nondominated point, implying that there is only one nondominated point. In other words, the ideal point, where each objective attains its optimal value, is feasible. In the following, we assume that (\(|\Y_N|>1\)). All $N$ extreme points and precisely one corresponding extreme flow can be determined by using the enhanced parametric programming approach in $\mathcal{O}(M+ Nn(m+n\log n))$ time~\citep{raith17}, where $M$ denotes the time required to solve a given single-objective minimum cost flow problem. Also, the algorithm stores one extreme flow for each extreme nondominated point.  

In~\cite{konen2023outputpolynomial}, it is shown that for each  maximal nondominated facet, we can create a reduced single-objective integer flow problem in which each image of a feasible flow lies on the maximal nondominated face. If we assume that we can solve the ACVF in output-polynomial time, we could determine for a maximal nondominated face all supported nondominated vectors  whose images lie on this facet in output-polynomial time. Thus, we solve the ACVF on the reduced network with only the first objective function. Solving the ACVF successively for the reduced single-objective minimum cost integer flow problems for each $N-1$ maximal nondominated facets would yield all supported nondominated vectors for the given BMCIF. Note that some vectors may lie in more than one maximal nondominated facet, namely the extreme nondominated vectors, and therefore, we have to ensure that we only store these flows once. In total, we would obtain all supported nondominated vectors in output-polynomial time. 
\end{proof}

\begin{theorem}\label{thm:all}
    If we can determine all supported nondominated vectors for a bi-objective minimum cost integer flow problem in output-polynomial time, then we can solve the ACVF in output-polynomial time. 
\end{theorem}

\begin{proof}
    Consider an instance of the ACVF. Construct a BMCIF by setting the objective as \( (c(f), -c(f)) \) on the same instance. That is, for each arc \( a \in A \), assign the cost vector \( (c_a, -c_a) \). Any distinct flow vector remains distinct, is nondominated, and lies on the hyperplane \( \{y \mid y_1 + y_2 = 0\} \). If we could determine all nondominated supported vectors in output-polynomial time, we could also  solve the ACVF for the original single-objective MCF.
\end{proof}

Let ACFV$^{\mathrm{Fin}}$ denote the decision problem of ACFV, i.e., given a set of distinct cost values \(\{c(f_1),\ldots,c(f_k)\}\) of some feasible flows, decide if there exists a feasible flow \(f_p\) with a distinct cost value to the given ones \(c(f_p)\notin\{c(f_1),\ldots,c(f_k)\}\).
Suppose we can show that ACFV$^{\mathrm{Fin}}$ is not solvable in polynomial time. In that case, there does not exist an output-polynomial time algorithm to determine all supported nondominated vectors unless $\mathbf{P} = \mathbf{NP}$.

Using these results, we can show that if we have given two distinct supported nondominated vectors $y^1,y^2$ with $y^1_1 < y^2_2$ for a bi-objective minimum cost integer flow problem (BMCIF), it is \textbf{NP}-hard to decide if  another supported nondominated vector $y^3$ exists, which has higher cost in the first objective for $y^1$ and lower cost in the first objective for $y^2$.

\begin{theorem} \label{thm:y3}
        Given two different supported nondominated vectors $y^1$ and $y^2$ of a BMCIF with $y^1_1< y^2_1$, the problem of deciding whether another supported nondominated vector $y^3$ with $y^1_1<y^3_1<y^2_1$ exists, i.e., in between the supported nondominated vector $y^1$ and $y^2$ is $\mathbf{NP}$-complete.
\end{theorem}

\begin{proof}

    The NP-completeness of the problem is established by reduction from the subset sum problem. Consider an instance of the subset sum problem and construct the single-objective minimum cost integer flow problem (MCIF) as in the proof of Theorem 3.11, with the corresponding construction shown in Figure 3.2. An optimal flow for this instance has a cost of $0$, while the highest cost of a feasible flow is \( \sum_{i \in N} w_i \). Transforming the single-objective MCIF into a BMCIF, as done in the proof of Theorem 3.15, ensures that all nondominated vectors are supported nondominated vectors and lie on the same edge between the two extreme supported points \( (0, 0)^\top \) and \( \left(\sum_{i \in N} w_i, -\sum_{i \in N} w_i\right)^\top \).  Introduce two arcs \( (s, t) \), one with cost \( (T - 1, -(T - 1)) \) and the other with \( (T + 1, -(T + 1)) \). This construction results in the two supported nondominated vectors \( y_1 = (T - 1, -(T - 1)) \) and \( y_2 = (T + 1, -(T + 1)) \). Any supported nondominated vector \( y_3 \) satisfying \( y_1^1 < y_3^1 < y_2^1 \) would have a cost equal to \( T \). Determining the existence of such a \( y_3 \) in polynomial time would allow solving the subset sum problem in polynomial time, leading to a contradiction.  
\end{proof}

\Cref{thm:y3} also implies  that it is $\textbf{NP}$-hard to determine the next best supported nondominated vector regarding the first or second objective for a given supported nondominated vector of a BMCIF. This means it is impossible to determine all supported nondominated vectors in a lexicographically ordered way regarding the first (second) objective in output-polynomial time. However, this theorem is insufficient to show that  the supported nondominated vector could not be obtained in output-polynomial time, and it remains an open question.

Consider the BMCIF of a subset sum instance in the proof of~\Cref{thm:y3}. We have shown that given a supported nondominated vector, it is  $\textbf{NP}$-hard to determine the next best supported nondominated vector for this instance. However, for this  instance of a BMCIF, we can determine all supported-nondominated vectors in output-polynomial time using a similar algorithm to the pseudo-polynomial labeling algorithm to solve the subset sum problem presented in \cite{PISINGER19991}.

\section{Determining the Supported Nondominated Vectors}
\label{sec:algo}
In this section, we briefly describe~\Cref{algo:allSupportedEfficientFlows}, presented in \cite{konen2023outputpolynomial}, which determines all supported efficient solutions for BMCIF. Additionally, an adjusted algorithm is derived to determine supported nondominated vectors  for BMCIF more efficiently when the required number of branches is significantly smaller than the number of supported efficient solutions. Note that~\Cref{algo:allSupportedEfficientFlows} is also presented for more than two objectives. However, this paper focuses on the bi-objective case.
\subsection{Determining All Supported Efficient Solutions}\label{chap:effsol}
\Cref{algo:allSupportedEfficientFlows} consists of two phases and relies on the following widely known fact for any integer supported flow, see, e.g., \cite{gal77}.

\begin{theorem}\label{thm:supp_face}
A flow $f$ is contained in $F_\X$, i.\,e., its image of $c(f)$ lies on a maximally nondominated face $F_\Y$ w.\,r.\,t.\ an associated weight vector $\lambda \in \Lambda_d$, if $f$ is an optimal solution to the parametric network flow program $P_\lambda$.
\end{theorem}

Any image of a \strictly supported flow must lie in at least one maximally nondominated face, and any integer point in a maximally nondominated face corresponds to a \strictly supported integer flow. 
Assuming that for each maximally nondominated face $F_i\in\{F_1,\dots, F_t\}$, one optimal solution $f^i$ and the corresponding weight vectors $\lambda^i$ are given, the problem of determining all \strictly supported flows reduces to determining all optimal flows for each parametric single-objective problem (weighted sum scalarization) $(P_{\lambda^i})$. These optimal solutions can be determined by  the algorithm for determining all optimal flows for single-objective minimum cost integer flow problems presented in~\cite{KONEN2022333}, which we refer to as the all optimal flows (AOF) algorithm. 
 The AOF algorithm successively searches for so-called \emph{proper zero-cost} cycles efficiently  using a modified depth-first search technique. 

 \begin{definition}[\citealt{Hamacher1995}]
	A \emph{proper $(i,j)$-cycle } $C\in D_f$ is a proper cycle that contains the arc $(i,j)$. In addition, a proper  $(i,j)$-cycle $C\in D_f$ is called \emph{minimal} if it has minimal cost among all proper ($i,j)$-cycles. 
\end{definition}

\begin{figure}[htb]
    \centering\small
    \begin{tikzpicture}[]
        \tdplotsetmaincoords{0}{0}
        \tikzstyle{vertex}=[circle,fill=black,draw=black,minimum size=4pt,inner sep=0]
        \tikzstyle{vertex2}=[circle,draw=DeepPink4,minimum size=4pt,inner sep = 0]
        \tikzstyle{vertex3}=[rectangle,fill=DodgerBlue4,draw=DodgerBlue4,minimum size=4pt,inner sep = 0]
        \tikzstyle{N_point}=[draw, cross out,scale=.5,thick]

        \begin{scope}[tdplot_main_coords, scale=.4]
          
        \draw [draw= white, fill= black!10  ,fill opacity=1 ]  
        (2,12.25) -- (2,9) -- (3,6)  -- (8,3) -- (12.5,3) -- (12.5,12.25)-- cycle;
	
        \draw[] (-1,0) -- (10,0);
        \draw[] (0,-1) -- (0,10);
    	\draw[->] (10,0) node[anchor=north]{} -- (13,0) node[anchor=north]{{$c^1(f)$}};
    	\draw[->] (0,10) -- (0,13) node[anchor=north east]{{$c^2(f)$}};

        \draw[thick] (2,9) -- (3,6) -- (8,3);
        \draw[dotted,thick] (2,9) -- (2,12.25);
        \draw[dotted,thick] (8,3) -- (12.5,3);
         
        \node[] at ( 8.5 ,11 ){\small $ \conv(\Y+\R^2_{\geqq}) $};

        \node[vertex3, label={[label distance=-5pt]above right:{}}] at (2,9) {};
        \node[vertex3, label={[label distance=-5pt]above right:{$y^i$}}] at (3,6) {};
        \node[vertex, label={[label distance=-5pt]above right:{}}] at (6,5) {};
        \node[vertex3, label={[label distance=-5pt]above right:{$y^{i+1}$}}] at (8,3) {};
        \node[vertex2, label={[label distance=-5pt]above right:{}}] at (4,5.4) {};
        \node[vertex2, label={[label distance=-5pt]above right:{}}] at (7,3.6) {};
        \node[vertex2, label={[label distance=-5pt]above right:{}}] at (2.5,7.5) {};
        \node[N_point, label={[label distance=-5pt]above right:{}}] at (3,9) {};
        \node[N_point, label={[label distance=-5pt]above right:{}}] at (7,7) {};

        \draw[black!50] (1.33,7) -- (9.67,2);
        \draw[thick,DodgerBlue4] (3,6) -- (8,3);

        \draw (9.7,1.8) node [anchor= east]    {$(\lambda^i)^\top C$};
        \draw[DodgerBlue4] (5.5,4) node [anchor= east]    {$F_i$};
        \end{scope}
        
\end{tikzpicture}
 \caption{Illustration of two consecutive extreme points, the maximally nondominated edge $F_i$ in blue and the cost vector of $(\lambda^i)^\top C$. }\label{Fig: segmentOnConvexHull}
\end{figure}
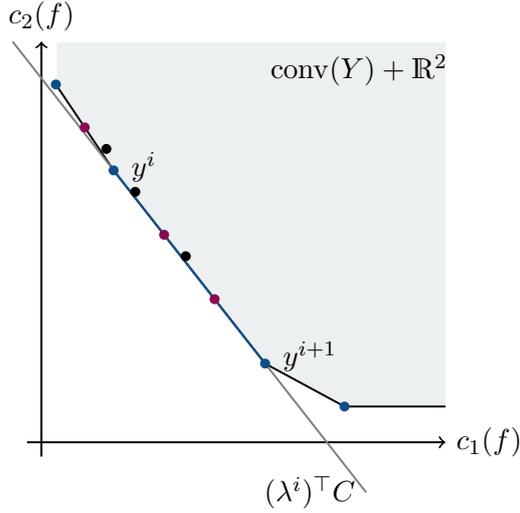

In order to determine all optimal solutions for a given maximal nondominated face, i.e., an edge between two consecutive extreme vectors $y^i$ and $y^{i+1}$ (see~\Cref{Fig: segmentOnConvexHull}), the approach uses the AOF algorithm on the network with weight vector $\lambda$ corresponding to the given edge $F_i$. The extreme vector $y_i$ is optimal for this network with cost function $\lambda^\top Cf$. In the first step, the network $D_{\lambda}$ gets reduced to $D^{'}_{\lambda}$ in the following way.

We remove all arcs in $\X \coloneqq \{(i,j)\in A\colon \overline{c}_{ij}\neq 0 \}$, i.e., arcs with non-zero reduced cost. Note that some of these arcs in $\X$ may carry flow. 
Consequently, we must adjust the demand vector accordingly to preserve feasibility. 
Consider the reduced network $D'_{\lambda}=(V,A^{\prime})$ with $A^{\prime}=A\backslash \X$ and the new flow balance values $b'_i = b_i - \sum_{(i,j) \in \X } f_{ij}+ \sum_{(j,i) \in \X} f_{ji}$. 
For any feasible flow $f$ in $D$, we can easily determine the corresponding feasible flow $f'$ in $D'$ where all flow values are copied for all arcs $a\in A^{\prime}$.

We denote by $\mathcal{P}^{\prime}_{\lambda}$ the set of all feasible integer flows of the reduced network $D^{\prime}_{\lambda}$. 
The fact that there is only a finite number of flows in $\mathcal{P}^{\prime}_{\lambda}$ follows from the assumption that there is no arc $(i,j)$ with $u_{ij}=\infty$. 

For a feasible solution $f' \in \mathcal{P'}_{\lambda}$, we define
$$ f^*(a)= \left\{ \begin{aligned}
	& f^{\prime}(a), & & \text{ if } a\in A^{\prime}, \\ &
	f(a), & &  \text{ if } a\in \X.
\end{aligned} \right.$$ 

The next lemma demonstrates how every optimal flow in $D_{\lambda}$ can be obtained from combining an initial optimal flow and some feasible flow in $D'_{\lambda}$. \\

\begin{lemma}[\citealt{KONEN2022333}] \label{lem:feasibleInD'=optimalInD}
	Let $f$ be an optimal integer solution of the MCIF in \(D_\lambda\). If $f^{\prime}$ is a feasible integer solution in $D^{\prime}_{\lambda}$ and $f^*$ is defined as above, then $f^*$ is an optimal solution in $D_{\lambda}$ and vice versa.
\end{lemma}

Given a feasible flow $f$ in $D'_{\lambda}$, we can compute another feasible flow by finding a proper cycle in the residual network of $D'_{\lambda}$. Therefore, we can determine another optimal flow in $D_{\lambda}$ by finding a proper cycle in $D'_{\lambda,f}$ instead of finding a proper $0$-cycle in $D_{\lambda,f}$. This gives us the following theorem:

\begin{theorem}[\citealt{KONEN2022333}]\label{theo:cycle->optimalflow}
	Let $f$ be an optimal integer flow. Then, every proper cycle in $D'_{\lambda,f}$ induces another optimal integer flow in $D_{\lambda}$.
\end{theorem}

Following this, the feasible outcome space is reduced to all feasible solutions whose image lies on the  edge $F_i$.  In~\cite{KONEN2022333}, it is shown that such a minimal proper cycle $C$ in $D'_{\lambda,f}$ can be determined in $\mathcal{O}(mn)$, yielding a new efficient solution whose image is on the maximal nondominated face $F_i$.  Since we can compute the next supported efficient solution in $D'_{\lambda}$, we can now apply the binary partition approach as used in~\cite{Hamacher1995} or~\cite{KONEN2022333}. 
Let $\X_{D'}$ be the set of optimal flows in $D'_{\lambda}$, which we also call the \emph{optimal solution space}, and let $f',f'' \in \X_{D'}$  as above. We then can recursively divide $\X_{D'}$ into $\X_{D'_1}$ and $\X_{D'_2}$ such that $f'$ and $f''$ are optimal flows in $\X_{D_1}$ and $\X_{D_2}$, respectively, and then seek more flows in $\X_{D'_1}$ and $\X_{D'_2}$. 
This can easily be done by identifying some arc $a$ with $f'(a)\neq f''(a)$. Such an arc must exist since $c(f')\neq c(f'')$. Let $(i,j) = a $ if $a\in L$ and $(i,j) = a^{-1}$ if $a\in U$.  
Assume w.l.o.g. that $f'_{ij} < f''_{ij}$. We set  
$$\X_{D'_1}\coloneqq \{z\in \X_{D'}: z_{ij}\le f'_{ij}\}$$ and $$\X_{D'_2}\coloneqq \{z\in \X_{D'}: z_{ij}\ge f'_{ij}+1\}.$$

This partitioning implies that each new minimum cost integer flow problem is defined in a modified flow network with an altered upper or lower capacity of a single arc $(i,j)$. By dividing the new solution spaces again until no other optimal flow exists, all optimal solutions are found. Therefore, all supported efficient solutions whose images lie on the maximal nondominated face $F_i$. We obtain the complete set of supported efficient solutions by doing so for each maximally nondominated face and only storing newly seen flows.  For each pair of consecutive extreme points $y^i$ and $y^{i+1}$, we determine the weight vector \(\lambda^i\in\Lambda\) that corresponds to the normal of the maximally nondominated facet \(F_i\) connecting the extreme points $y^i$ and $y^{i+1}$:

\begin{align*}
  \lambda^i \coloneqq 
  \begin{pmatrix}
      c^2(f^i)-c^2(f^{i+1})\\
      c^1(f^{i+1})-c^1(f^i)
  \end{pmatrix}
\end{align*}

Then $f^i$ and $f^{i+1}$ are both optimal flows for the single-objective weighted sum scalarization $(P_{\lambda^i})$~\citep{Eusebio09}. Hence, determining all optimal solutions for $(P_{\lambda^i})$ gives all supported efficient flows whose image lies in between $F_i$.

\begin{algorithm}[htb]
\KwData{ $(D,l,u,b,c^1,c^2)$ }
\KwResult{ The  set of all supported flows $\X_S$}
\(\X_S\leftarrow \varnothing\)\;
\tcp{Determine all extreme supported points~$y^i$ and for each one corresponding extreme flow $f^i$, sorted non-decreasingly in \(c^1(f)\)}
$\{(y^i,f^i) \colon i\in \{1,\dots,N\}\}\leftarrow $ EnhancedParametricNetworkAlgortihm(D) \;
\For{$i=1,\dots,N-1$}{
    \(\lambda_1^i \leftarrow c^2(f^i)-c^2(f^{i+1}) \);\quad  \(\lambda_2^i \leftarrow c^1(f^{i+1})-c^1(f^i) \) \; 
    $\bar{c} \leftarrow$ Determine reduced costs\;
    $\X_{\lambda^i} \leftarrow \{f^i\} $ \; 
	\(\X_S\leftarrow \X_S\, \cup\,\)FindAllOptimalFlows($D,\bar{c},\lambda^i,f^i, \X_{\lambda^i}$) \;
	\tcp{Return only flows with $c^1(f)\neq c^1(f^i)$ to avoid repetitions.}
}
\BlankLine
\caption{FindAllSupportedEfficientFlowsBiObjective}\label{algo:allSupportedEfficientFlows}
\end{algorithm}

\begin{algorithm}[htb]
	\KwData{optimal flow $f$ for $P_{\lambda^i}$, the reduced costs $\overline{c}$, the set of all current found   integer flows $\X_{\lambda^i}$ on $P_{\lambda^i}$, $D$}
	\KwResult{The set of optimal integer flows $\X_{\lambda^i}$ for $P_{\lambda^i}$}
	\BlankLine

	$f^* \leftarrow$ FindAnotherOptimalFlow($f$, $\overline{c}$, $D$)\;
	\lIf{$f^*=$ \Null}{\KwRet $\X_{\lambda^i}$}
	\BlankLine
	$\X_{\lambda^i} \leftarrow \X_{\lambda^i} \cup (f^*)$\;
	\tcp{Partition solution space and find new optimal flow}
	$a \leftarrow$ arc $a$ with $f(a) \neq f^*(a)$\;
	\eIf{$f(a) < f^*(a)$}
	{ $\X_{\lambda^i} \leftarrow \X_{\lambda^i} \, \cup \,$ \{FindAllOptimalFlows($D,\bar{c},\lambda^i,f, \X_{\lambda^i}$) with $u_a=f(a)$\}\;
	$\X_{\lambda^i} \leftarrow \X_{\lambda^i} \, \cup \,$	\{FindAllOptimalFlows($D,\bar{c},\lambda^i,f^*, \X_{\lambda^i}$) with $l_a=f(a)+1$\}\;
	}
	{ $\X_{\lambda^i} \leftarrow \X_{\lambda^i} \, \cup \,$\{FindAllOptimalFlows($D,\bar{c},\lambda^i,f, \X_{\lambda^i}$) with $l_a=f(a)$\}\;
	$\X_{\lambda^i} \leftarrow \X_{\lambda^i} \, \cup \,$	\{FindAllOptimalFlows($D,\bar{c},\lambda^i,f^*, \X_{\lambda^i}$) with $u_a=f(a)-1$\}\;
	}	
	\caption{FindAllOptimalFlows}\label{algo:FindAllOptimalFlows}
\end{algorithm}

\subsection{The Adjusted Algorithm}\label{sec:adj}

In the current version of the algorithm, we search for a proper cycle in the reduced network to obtain a second supported efficient solution. Note that in the reduced network, any cycle would have a cost equal to zero regarding the weighted sum scalarization with a weight vector $\lambda$ corresponding to the current maximal nondominated face $F_i$. However, we will now consider the first objective. We then search for a minimal proper cycle in the reduced network, i.e., a proper cycle with minimum cost regarding the first objective under all proper cycles in $D'_\lambda$. A proper zero cost cycle regarding the first objective in $D'_\lambda$ would yield a supported efficient solution whose images map to the same vector as the initial solution (in the bi-objective case). Since we want, in the best case, only one flow per vector, we search for a minimal proper cost cycle under all proper cycles that have costs greater than zero. We then would obtain a distinct cost second-best flow regarding the first objective. Using the complementary slackness condition, we prove that we could obtain such a cycle in $\mathcal{O}(n^3)$. Using this fact, we could adjust the previously presented algorithm in~\Cref{chap:effsol} to find the supported nondominated vectors more efficiently.

 We want to find all supported nondominated vectors for each maximally face $F_i$ (in this case, given by an edge between two consecutive extreme points). For that, we want to determine all supported nondominated vectors between the edge of the two extreme vectors $y^i$ and $y^{i+1}$. In~\Cref{Fig: segmentOnConvexHull}, an example is shown. 

 For $y_i$, let $f'$ be a corresponding optimal flow. Considering the edge $F_i$, we know that $f'$ is not only an optimal flow for the weight vector $\lambda$, but as well $f'$ is an optimal flow regarding the first objective $c^1$ in $D'_{\lambda}$.  So for $f'$, there cannot be a negative cycle regarding the cost of $c_1$ in $D'_{\lambda,f'}$ and the complementary slackness conditions hold. 

In order to find the next nondominated vector on the edge $F_i$, we determine the minimal proper cycle $C\coloneqq  \operatorname{argmin} \{ c^1(C) \colon c^1(C)>0, \;C\in D'_{\lambda,f'}\} $. Using techniques from \cite{KONEN2022333} and \cite{Hamacher1995} this can be done in $\mathcal{O}(n^3)$. 
Note that this is feasible only for the first and last supported nondominated vector on a maximally nondominated face since it requires an optimal solution w.r.t.\ the first or second objective function to start with. Thus, this procedure cannot  be extended to iteratively generate supported nondominated vectors along the face in an ordered way.

Let $\overline{D}=(\overline{d}_{ji})$ be the distance table of $D_{\lambda,f}$ concerning $\overline{c}^1(f)$, i.e., $\overline{d}_{ji}$ is the length of the shortest path $P_{ji}$ from $j$ to $i$ in $D_f$ with length $\overline{c}^1(f,P_{ji})$. 
The distance table may be computed in time $\mathcal{O}(n^3)$ using the Floyd-Warshall algorithm or in time $\mathcal{O}(n^2\log n+mn)$ by (essentially) repeated calls to Dijkstra's algorithm; the latter is more efficient on sparse graphs~\cite{Hamacher1995}.

Hamacher shows the following property for proper minimal $(i,j)$-cycles:

\begin{property}[\citealt{Hamacher1995}]\label{min-cylce}
	For any anti-parallel arc in $A_f$, i.e., $(i,j)\in A_f$ with  $(j,i)\notin A_f$, the cost of a proper minimal $(i,j)$-cycle $C\in D_f$ is given by $c(f,C)=\overline{c}_{ij}(f)+\overline{d}_{ji}$.
\end{property} 

When using Hamacher's idea, the problem is that it only applies to arcs with no anti-parallel arc in $A_f$. For all other arcs, the cost of the corresponding proper minimal $(i,j)$-cycle has to be computed by finding a shortest path in $D_f \setminus \{(j,i)\}$.
In the following result, we show that we can use the complementary slackness optimality conditions of an optimal solution to overcome this problem and restrict ourselves to consider only arcs with no anti-parallel arcs in $D_{\lambda,f}$. 

\begin{theorem}
	We can determine the minimal cost cycle 
    \[ C\coloneqq \argmin_{C\in D'_{\lambda,f'}} \{ c^1(C) \colon c^1(C)>0\} \]
    over all cycles with strictly positive weight in $\mathcal{O}(n^3)$.
\end{theorem}
\begin{proof}
	Let $C$ be a minimal proper cycle with the property $c^1(C)\neq 0$. 
	So $\overline{c}^1(C,f)>0$, because	there cannot be a cycle with negative costs due to the negative cycle optimality conditions. 
	Therefore, there exists an arc $(i,j)\in C$ with $\overline{c}_{ij}(f)>0$. Since $f$ is an optimal solution, the complementary slackness optimality conditions ensure that $\overline{c}^1_{ij}(f)> 0$ only holds if $(i,j)\in A $ and $f_{ij}=l_{ij}$ or $(j,i)\in D$ and $f_{ji}=u_{ji}$. 
	Therefore, we have that $(i,j)\in \overline{A}\coloneqq \{(i,j)\in A_f \colon  (i,j)\in A \text{ with }  f_{ij}=l_{ij} \text{ or } (j,i)\in A \text{ with }  f_{ji}=u_{ji}\}$ and $(j,i)\notin D_f$, (since the flow value of the corresponding arc of $(i,j)$ in $D$ is equal to the upper or lower capacity of this arc). 
 
	Since $(j,i)\notin D_f$ it holds that $\overline{c}^1(C,f)= \overline{c}_{ij}+ \overline{d}_{ji}$ due to~\Cref{min-cylce}. Since $\overline{c}^1_{ij}(f) \geq 0$ any cycle $C$ with $c^1(C)$ has at least one of such an arc. Consequently, we can determine a minimal proper cycle  $C\coloneqq  \operatorname{argmin} \{ c^1(C) \colon c^1(C)>0,\; C\in D'_{\lambda,f'}\}$ by just choosing
	$C=\argmin\{c(f,C)=\overline{c}_{ij} + \overline{d}_{ji} \colon  C=\{(i,j)\}\cup P_{ji} \text{ with } (i,j)\in \overline{A}\} $.
    We can compute the distance table in $\mathcal{O}(n^3)$ time.  Notice that this approach also provides the paths $P_{ji}$ in addition to the cost $\overline{d}_{ji}$.
	
	Given all pairwise distances $\overline{d}_{ji}$ and the reduced costs $\overline{c}^1_{ij}(f)$ for all $(i,j)$ in $D^1_f$ we can determine $c^1(f,C)=\overline{c}^1_{ij} + \overline{d}_{ji}$ in constant time $\mathcal{O}(1)$ and can determine the $\argmin$ in $\mathcal{O}(m)$ time. As a result, we can compute a minimal proper cycle with $c^1(C)\neq 0 $ in time $\mathcal{O}(n^3)$. 
\end{proof}

So let $C$ be chosen as above. 
Let $f''= f' + \chi(C)$. The flow $f''$ is  supported, and its image lies on the edge $F_i$. We are going to prove that $f''$ is the next nondominated point on $F_i$ regarding the cost of $c^1$, i.e., there does not exist a flow $\bar{f}$ whose image lies on $F_i$ with the property $c^1(f') < c^1(\bar{f}) < c^1(f'')$.

\begin{theorem}\label{theo: no flow}
	There exists no flow $\bar{f}$ whose image \(c^1(\bar{f})\) lies on $F_1$ between the images of \(f'\) and \(f''\), i.e., $c^1(f') < c^1(\bar{f}) < c^1(f'')$.
\end{theorem}  

\begin{proof}
 It holds that 
 \begin{align*}
    &f'' = f'+ \chi(C) \qquad \text{and thus}\quad c^1(f'') = c^1(f') + c^1(C)
    \intertext{and we know that any flow $\bar{f}$ on $F_1$ can be written as}
    &\bar{f} = f' + \sum_{C_i \in D'_{\lambda,f'}} \chi(C_i)  \qquad c^1(\bar{f}) = c^1(f') + \sum_{C_i \in D'_{\lambda,f'}} c^1(C_i).\\
    \implies\; & c^1(f') < c^1(f')+ c^1(C) = c^1(f'') \leq c^1(f')+ \sum_{C_i \in D'_{\lambda,f'}} c^1(C_i) = c(\bar{f})  .
 \end{align*}
 The inequality $\sum_{C_i \in D'_{\lambda,f'}} c^1(C_i) \geq c^1(C)$ holds since the sum contains the positive cost of at least one cycle regarding the first objective. Since $C$ was chosen as the minimal cost cycle
 $$C\coloneqq  \operatorname{argmin}_{C\in D'_{\lambda,f'}} \{ c^1(C) \colon c^1(C)>0\}, $$
 it holds that $c^1(f'') \leq c^1(\bar{f})$. 
\end{proof}

Since we can compute the next nondominated supported vector in $D'_{\lambda}$, we can now apply the binary partition approach as used in~\cite{Hamacher1995} and \cite{KONEN2022333}, presented above. Thereby, we substitute $\operatorname{FindAnotherOptimalFlow}$ in~\Cref{algo:FindAllOptimalFlows} by~\Cref{algo:SecondBest}.
\begin{algorithm}[htb]
	\KwData{optimal flow $f$ in $P_{\lambda^i}$}
	\KwResult{A second distinct cost-best flow $f^{\prime}$ with $c(f^{\prime})\neq c'(f) $, if one exists.}
	\BlankLine
	
	\BlankLine
	$y$ $\leftarrow$ ComputeNodePotential($f$, $D$)\;
	$\overline{c}$ $\leftarrow$ ComputeReducedCost($y$, $D$)\;

		$(\overline{D},P) \leftarrow$ DetermineDistanceTableAndPaths($\overline{c},f,D$)\;
		$C\leftarrow\argmin\{c(f,C)\}= \argmin_{C\in D'_{\lambda,f'}} \{ c^1(C) \colon c^1(C)>0\} $\;
	 \lIf{$C = \Null$} {\KwRet $\emptyset$}
		$f'\leftarrow f + \chi(C)$\;
	\BlankLine
	\KwRet{$f'$\;}
	\caption{FindSecondDistinctCostBestFlow}\label{algo:SecondBest}
\end{algorithm}

This adjusted algorithm can be more efficient in determining all supported nondominated vectors than by determining all supported efficient flows, as the next example will show, since we might have an exponential number of supported efficient flows that we would not consider with this new technique. 

\begin{example}\label{ex:1}
    Consider a graph $D$ with nodes $\{1,\ldots, k\}$, parameters $L,M\in\N$ and an optimal flow $f$ w.r.t.\ the first objective. Node~$1$ has $b_1=2$ and node~$k$ has $b_k=-2$. For all other nodes  $i\in \{2,\ldots, k-1\}$ we have $b_i=0$. The graph $D$ contains arcs 
    \begin{itemize}
        \item $(i,i+1)$ for all $i\in \{1,\ldots, k-1\}$ with $u_{i,i+1} = (k-3)\,M+(L+2)$, $f_{i,i+1} = 2$ and reduced costs  $\bar{c}^1_{i,i+1} =\bar{c}^2_{i,i+1}=0$,
        \item arcs $(k-i,k-i-2)$ for all $i\in \{1,\ldots k-3 \}$ with $u_{k-i,k-i-2} =M$, $f_{k-i,k-i-2}=0 $ and  $\bar{c}^1_{k-i,k-i-2} =\bar{c}^2_{k-i,k-i-2}=0$, and
        \item one arc $(k,k-2)$ with $u_{k,k-2} =L$, $f_{k,k-2}=0 $ and  $\bar{c}^1_{k,k-2}=1,   \bar{c}^2_{k,k-2}=-1$. 
    \end{itemize}  
    \Cref{example1} illustrates this example for $k=5$. There are $L+1$ nondominated vectors, which are all supported. However, we would have $(M+1)^{k-3}(L+1)$ efficient flows, which are all supported. However, using the branching technique described above, we would have only $L+1$ leaves at the end of our branching.

\end{example}

\tikzstyle{vertex}=[circle,fill=black,draw=black,minimum size=8pt,inner sep=0]%
\tikzstyle{edge} = [draw,thick,->]%
\tikzstyle{selected vertex} = [vertex, fill=red!24]%
\tikzstyle{selected edge} = [draw,line width=5pt,-,yellow!50]%
\begin{figure}[htb]
	\footnotesize
	\centering
	\begin{tikzpicture}[ ->,shorten >=1pt,auto,node distance=3cm,%
		semithick]%
		\foreach \pos/\name/\label in {{(-2,0)/a/2}, {(-0,0)/b/}, {(2,0)/c/},%
			{(4,0)/d/}, {(6,0)/e/-2}}%
		\node[vertex] (\name) [label=$\label$] at \pos {};%
		\path (a) edge[->,thick]   	 (b)%
		(b) edge[->,thick]   	 (c)%
		(c) edge[->,thick]    (d)%
		(d) edge[->,thick]   	 (e)%
		(d) edge[->,thick,bend right]  node[sloped] {$(M,0,0,0)$} (b)
		(c) edge[->,thick,bend right]  node[sloped] {$(M,0,0,0)$} (a)
		(e) edge[->,thick,bend left]  node[sloped,below] {$(L,0,1,-1)$} (c);
	\end{tikzpicture}%
	\caption{Graph $D$ of the BMCIF in~\Cref{ex:1} with an optimal flow $f$ w.r.t.\ the first objective. The arcs are labeled with $(u_a,f_a,\bar{c}^1_{ij},\bar{c}^2_{ij}).$ Here  $l_{ij}=0$  for all arcs and  $(u_a=(k-3)M+(L+2),f_a=2,\bar{c}^1_{ij}=0,\bar{c}^2_{ij}=0)$ for all arcs that are not labeled. The nodes are labeled with $b_i$ and $b_i=0$ for all nodes which are not labeled. $M> L$.}\label{example1}
\end{figure}
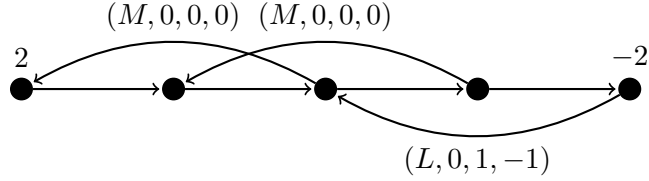

However, as shown in the following example, we might find supported efficient flows (even exponentially many) with images corresponding to supported nondominated vectors already found in other branches.
\begin{example}\label{ex:2}
Consider a graph $D$ with nodes $\{1,\ldots, k\}$, parameter $L\in\N$ and an optimal flow $f$ w.r.t.\ the first objective. Node~$1$ has $b_1=2$ and node~$k$ has $b_k=-2$. All other nodes~$i\in \{2,\ldots, k-1\}$ are transshipment nodes, $b_i=0$. The graph $D$ contains arcs 
\begin{itemize}
    \item $(i,i+1)$ for all $i\in \{1,\ldots, k-1\}$ with $u_{i,i+1} = L+2$, $f_{i,i+1} = 2$ and reduced costs  $\bar{c}^1_{i,i+1}=\bar{c}^2_{i,i+1}=0$, and
    \item $(k,k-i)$ for all $i\in \{2,\ldots,k-1\}$ with $u_{k,k-1} = L$, $f_{k,k-1} = 0$ and reduced costs  $\bar{c}^1_{k,k-1}=1,\bar{c}^2_{k,k-i}=-1$.
\end{itemize}
\Cref{example2} illustrates the graph $D$ for $k=5$. Then, this BMCIF has $L+1$ nondominated vectors, and all of them are supported. However, we have $\sum_{i=0}^{L} \binom{(k-2)+i-1}{i}$ supported efficient flows, and all of them would yield a leave of our branching technique. 
This number can be exponential. Assume that $L>k$, then it holds that 
\[
    \sum_{i=0}^{L} \binom{(k-2)+i-1}{i}  > \sum_{i=0}^{k} \binom{k-3+i}{i}  > \sum_{i=0}^{k} \binom{k-3}{i} = 2^{k-3}. 
\] 
Since $k\in \mathcal{O}(m)$, we have an exponential amount of flows to consider. 

\end{example}

\begin{figure}[htb]
	\centering
	\centering
	\begin{tikzpicture}[ ->,shorten >=1pt,auto,node distance=2.8cm,%
		semithick]%
		\foreach \pos/\name/\label in {{(-2,0)/a/2}, {(-0,0)/b/}, {(2,0)/c/},%
			{(4,0)/d/}, {(6,0)/e/-2}}%
		\node[vertex] (\name) [label=$\label$] at \pos {};%
		\path (a) edge[->,thick]   	 (b)%
		(b) edge[->,thick]   	 (c)%
		(c) edge[->,thick]    (d)%
		(d) edge[->,thick]   	 (e)%
		(e) edge[->,thick,bend left]  node[sloped] {\footnotesize$(L,0,1,-1)$} (c)
		(e) edge[->,thick,bend right]  node[sloped] {\footnotesize$(L,0,1,-1)$} (b)
		(e) edge[->,thick,bend left]  node[sloped,below] {\footnotesize$(L,0,1,-1)$} (a);
	\end{tikzpicture}%
	\caption{Graph $D$ of the BMCIF in~\Cref{ex:2} with an optimal flow $f$ w.r.t.\ the first objective. The arcs labeled with $(u_a,f_a,\bar{c}^1_{ij},\bar{c}^2_{ij}).$ Here  $l_{ij}=0$  for all arcs and  $(u_a=L+2,f_a=2,\bar{c}^1_{ij}=0,\bar{c}^2_{ij}=0)$ for all arcs that are not labeled. The nodes are labeled with $b_i$ and $b_i=0$ for all nodes which are not labeled.  }\label{example2}
\end{figure}

\subsection{$\varepsilon$-Constraint Scalarizations on the Reduced Networks}
One decision-space method for determining nondominated vectors for general multi-objective linear programs is the well-known $\varepsilon$-constraint method \citep{Haimes1971}. In the $\varepsilon$-constraint method, there is no
aggregation of criteria as in the weighted sum scalarization. Instead, only one of the original objectives is minimized
while the others are transformed into constraints. 

The $\varepsilon$-constraint scalarization of \eqref{eq:MMCIF} can be represented as:
\[
    \begin{array}{r@{\extracolsep{1.3ex}}r@{\extracolsep{.75ex}}c@{\extracolsep{.75ex}}l}
        \min & c^j(f)\\
        \text{s.t.} &  c^k(f)&\leq& \varepsilon_k \qquad \forall k\in\{1,\ldots,d\}, \; k\neq j\\
        & f&\in& \X
    \end{array}
\]
     
An illustration of the $\varepsilon$-constraint for a bi-objective minimum cost integer flow problem can be found in~\Cref{Fig: eps}.

\begin{figure}
    \centering\small
    \begin{tikzpicture}[]
        \tdplotsetmaincoords{0}{0}
        \tikzstyle{vertex}=[circle,fill=black,draw=black,minimum size=4pt,inner sep=0]
        \tikzstyle{vertex2}=[circle,draw=DeepPink4,minimum size=4pt,inner sep = 0]
        \tikzstyle{vertex3}=[rectangle,fill=DodgerBlue4,draw=DodgerBlue4,minimum size=4pt,inner sep = 0]
        \tikzstyle{N_point}=[draw, cross out,scale=.5,thick]

        \begin{scope}[tdplot_main_coords, scale=.4]
          
        \draw [draw= white, fill= black!10  ,fill opacity=1 ]  
        (2,12.25) -- (2,9) -- (3,6)  -- (8,3) -- (12.5,3) -- (12.5,12.25)-- cycle;
	
        \draw[] (-1,0) -- (10,0);
        \draw[] (0,-1) -- (0,10);
    	\draw[->] (10,0) node[anchor=north]{} -- (13,0) node[anchor=north]{{$c^1(f)$}};
    	\draw[->] (0,10) -- (0,13) node[anchor=north east]{{$c^2(f)$}};

        \draw[thick] (2,9) -- (3,6) -- (8,3);
        \draw[dotted,thick] (2,9) -- (2,12.25);
        \draw[dotted,thick] (8,3) -- (12.5,3);
         
        \node[] at ( 8.5 ,11 ){$ \conv(\Y+\R^2_{\geqq}) $};

        \node[vertex3, label={[label distance=-5pt]above right:{}}] at (2,9) {};
        \node[vertex3, label={[label distance=-5pt]above right:{}}] at (3,6) {};
        \node[vertex, label={[label distance=-5pt]above right:{}}] at (6,5) {};
        \node[vertex3, label={[label distance=-5pt]above right:{}}] at (8,3) {};
        \node[vertex2, label={[label distance=-5pt]above right:{}}] at (4,5.4) {};
        \node[vertex2, label={[label distance=-5pt]above right:{}}] at (7,3.6) {};
        \node[vertex2, label={[label distance=-5pt]above right:{}}] at (2.5,7.5) {};
        \node[N_point, label={[label distance=-5pt]above right:{}}] at (3,9) {};
        \node[N_point, label={[label distance=-5pt]above right:{}}] at (7,7) {};

        \draw[color=orange,thick,latex-latex] (-1,7.25) -- (-1,8.5) -- (13.5,8.5) -- (13.5,7.25);
	\node at (-1.5,8.5){$\varepsilon$};
        
        \end{scope}
        
\end{tikzpicture}
\caption{An example of the $\epsilon$-constraint scalarization  $ \{ \min_{f\in \X} c^1(f) \colon c^2(f)\leq \varepsilon \}$ for a BMCIF.}
\label{Fig: eps}
\end{figure}

For the BMCIF, all nondominated vectors can be found by solving a sequence of $\varepsilon$-constraint problems.  Starting with the lexicographically minimal solution \newline $\operatorname{lexmin}\{c^1(f),c^2(f)\}$ regarding the first objective, which can be determined in polynomial time, we can determine the next nondominated point using the $\varepsilon$-constraint. After each newly generated point, we update $\varepsilon$ until the last nondominated point $\operatorname{lexmin}\{{c^2(f),c^1(f)}\}$ is generated. In \cite{Eusebio09}, an implicit enumeration algorithm for BMCIF is given, which solves such a sequence of $\varepsilon$-constraint problems by computing optimal non-integer solutions
with a network simplex algorithm and then determining optimal integer solutions with a branch-and-bound technique. 

We also could use this $\varepsilon$-constraint method to obtain the complete set of supported nondominated vectors for the BMCIF, in which we have to solve one $\varepsilon$-constraint problem for each supported nondominated vector. Thereby, we solve the $\varepsilon$-constraint method on each maximally nondominated face. Let $X_{D'}$ be the set of all feasible flows for the reduced network for one maximal nondominated face, i.e., all solutions whose image lies on this maximal nondominated face. Any solution in the reduced network is a supported efficient solution. Using the $\varepsilon$-constraint method on all of these facets would determine all supported nondominated vectors. We only must solve one $\varepsilon$-constraint problem for each supported nondominated vector. Let $T(\varepsilon)$ be the longest time required to solve such a $\varepsilon$-constraint problem.

\begin{theorem}
   For a given BMCIF, we can determine all $S$ supported nondominated vectors in time $\mathcal{O}(N\, n(m + n log n) + M + S(T(\varepsilon))$.
\end{theorem}

\begin{proof}

The enhanced parametric network approach~\citep{raith09} requires $\mathcal{O}(N n(m+n \log n) + M)$ time,  
where $M$ is the time required to solve a single-objective minimum cost integer flow problem, and $N$ is the number of extreme supported nondominated vectors.  
Since the algorithm determines the extreme points in decreasing order of $c_1(f)$, no additional time is needed for sorting.  
Defining the weight vectors $\lambda^i$ for $i = 1, \dots, N-1$ and constructing the network with the corresponding cost function takes  
$\mathcal{O}(N (n + m))$ time. Determining all $S_i$ supported nondominated vectors on the maximally nondominated face $F_i$  
requires solving $S_i - 1$ $\varepsilon$-constraint problems. Doing so for each maximally nondominated face requires  
$\mathcal{O}(S T(\varepsilon))$ time.  
\end{proof}

\subsection{A more compact formulation for the ILP in the $\varepsilon$-method}
The $\varepsilon$-constraint problem contains $m$ variables and $2\,m+n+1$ constraints. However, according to~\Cref{theo:optimalFCompositionOfCa}, each flow $f^*$ can be written by an initial optimal tree solution $f$ and a conical combination of incidence vectors of all \emph{induced cycles} with bounded coefficients, i.e., 
	$$ f^* = f + \sum_{a \notin T} \lambda_a\, \chi(C_a)$$
	for some $\lambda \in \mathbb{Z}$ and it holds that 
        $$c(f^*)= c(f) + \sum_{a \notin T} \lambda_a\, c(C_a). $$

Therefore, instead of solving a constrained minimum cost integer flow problem, we could also solve the following ILP, which searches for the best combination of induced cycles such that the capacity constraints are satisfied. 
\begin{alignat*}{3}
        \min&&\quad& \sum_{a\notin T} \lambda_a\, c^1(C_a) \label{eq:cMMCIF}\tag{c-MMCIF}\\
        \text{s.t.}&&&  0-f_{ij} \leq \sum_ {a: (i,j) \in C_{a}} \lambda_a\,\chi_{ij}(C_{a}) \leq u_{ij} -f_{ij}&\quad& \forall (i,j) \in \bigcup_{a  \notin T} C_{a} \\
        & && \sum_{a\notin T} \lambda_a \, c^2(C_a) \leq \varepsilon
\end{alignat*}

\begin{theorem}
    For $\varepsilon$ sufficiently large, the set of feasible solutions of \eqref{eq:MMCIF} and \eqref{eq:cMMCIF} coincide. 
\end{theorem}

\begin{proof}
Let $f$ be an initial optimal tree solution. 
Assume that $f^*$ is a feasible solution for~\ref{eq:MMCIF}. After~\Cref{theo:optimalFCompositionOfCa}, the solution $f^*$ can be written as  $f^* = f + \sum_{a \notin T} \lambda_a^*\, \chi(C_a)$ for some $\lambda \in \mathbb{Z}$. Let $\lambda^*_a$ for $a\notin T$ be the solution for~\ref{eq:cMMCIF}, then  $0-f_{ij} \leq \sum_ {a: (i,j) \in C_{a}} \lambda_a\,\chi_{ij}(C_{a}) \leq u_{ij} -f_{ij}$ for all $(i,j) \in \bigcup_ {a  \notin T} C_{a}$ would be satisfied, since otherwise an arc  $(u,v)$ exists  where the capacity constraint $0\le f^*_{uv} \le u_{uv}$ would not hold. A contradiction of the feasibility of $f^*$.  
Now assume that $\lambda_a^{\prime}$ for $a\notin T$ is a feasible solution of~\ref{eq:cMMCIF}. Let $f^{\prime} = f + \sum_{a \notin T} \lambda_a'\, \chi(C_a)$. Since $f$ was an initial optimal tree solution and we only change flow on cycles, it holds that $\sum_{j:(i,j)\in A}f_{ij}' - \sum_{j:(j,i)\in A} f_{ji}' = b_i$ for all $i\in V$. The boundaries of $0 \le f_{ij} \le u_{ij}$ for all $(i,j)\in A$ are also satisfied, since otherwise  $u_{ij} -f_{ij}$ for one arc $(i,j) \in \bigcup_ {a  \notin T} C_{a}$ would not be satisfied. A contradiction.

\end{proof}

This ILP has $n$ variables and at least $n-1$ constraints less than the standard $\varepsilon$-constraint problem.
According to~\Cref{theo:optimalFCompositionOfCa}, the respective sets of feasible solutions coincide.  In~\Cref{chapt:Numerical}, we evaluate numerically which approach is faster in practice, the standard $\varepsilon$-constraint method or the combination approach of the induced cycles. We also compare the running times to determine all supported nondominated vectors compared to the algorithms in~\Cref{chap:effsol} and~\Cref{sec:adj}.

\section{Numerical experiments}\label{chapt:Numerical}
This section presents the implementation and numerical evaluation of the four methods. The section aims to report and compare the results, providing a comprehensive understanding of their behaviors.

All computations are conducted on a computer with an Intel\textregistered\ Core\texttrademark i8-8700U CPU 3.20 GHz processor with 32 GB RAM, using a LINUX operating system. The algorithms are implemented and run in Python (Version 3.11). In addition, for solving the $\varepsilon$-constraint scalarizations, Gurobi 12.1 embedded in Python is used. To ensure fair comparisons, the ILPs in Gurobi were solved using a single thread.

For  the computational experiments, we utilized test instances from~\Cref{example1} and~\Cref{example2}, as well as minimum cost integer flow problem classes generated by the NETGEN network generator~\citep{Klingman1974NETGENAP}. The entire test comprised $10$ problem classes, with each class consisting of a set of 15 randomly generated network problems.  The parameters that allowed the random generation of each NETGEN instance are the number of nodes, arcs, and nodes acting as supply or sink nodes, respectively, the greatest cost, greatest capacity, and the total supply in each network. In each problem class, the number of arcs and nodes varies.  The instances ranged from $50$ to $2000$ nodes with $100$ to $8000$ arcs. These variables were chosen as independent variables due to their direct influence on the number of possible supported nondominated vectors or supported efficient solutions and, therefore, influencing the different number of iterations of the different methods. All the other parameters that NETGEN can
accept were kept constant. All instances have two nodes acting as supply nodes and two as 
sink nodes, a maximum arc cost of 10 for both objective functions, a maximum upper capacity of 50, and a total supply of 50. 

Results for the NETGEN instance classes are presented in~\Cref{tab2}, while the results of the test instances from~\Cref{example1} and~\Cref{example2} are summarized in~\Cref{tab3} and~\Cref{tab4}.  These tables display the number of extreme nondominated vectors, supported nondominated vectors, supported efficient solutions, and CPU time for all four methods. We display the min, max, and mean of the times and numbers of the 15 network problems.  Note that we only consider problem instances with at least two nondominated vectors.

\begin{table}[htb]\footnotesize
\caption{Numerical results for the different instance classes generated with NETGEN. T-$|\Y_{EN}|$ displays the time needed to determine the number of extreme vectors. T-AO, T-DS, T-$\varepsilon$, T-New-$\varepsilon$ refers to the CPU time in seconds needed for~\Cref{algo:allSupportedEfficientFlows}, the adjusted version with~\Cref{algo:SecondBest}, the $\varepsilon$-Method, and the new more compact $\varepsilon$-Method. Empty entries (--) reflect a CPU time of over 500 seconds and have not been recorded.}\label{tab2}
\vspace{0.5cm}
\begin{tabular}{llrrrrrrrr}
\toprule
 Class &{} &  $|\Y_{EN}|$ &  T-$|\Y_{EN}|$ &  $|\Y_{SN}|$ &  $|\X_{SN}|$ &  T-AO &   T-DS &  T-$\varepsilon$&  T-New-$\varepsilon$ \\
\midrule
{1} &min  & 2  & 0.027  & 4  & 4  & 0.0058  & 0.1010  & 0.0049  & 0.0027  \\
$n=50$ & max  & 10 & 0.163  & 41 & 47  & 0.0887  & 1.1871  & 0.0287  & 0.0161  \\
$m=100$ & mean & 4.9  & 0.074  & 20.7  & 22.2  & 0.0269  & 0.5565  & 0.0131  & 0.0072  \\
\midrule
{2} &min  & 4  & 0.125  & 23  & 29  & 0.0408  & 0.7778  & 0.0156  & 0.0097  \\
$n=50$ & max  & 15  & 0.589  & 97  & 227  & 0.3906  & 8.0199  & 0.1056  & 0.0296  \\
$m=200$ & mean & 9.2  & 0.340  & 42.3  & 58.5  & 0.0984  & 1.8382  & 0.0399  & 0.0181  \\
\midrule
{3} &min  & 2  & 0.103  & 3  & 3  & 0.0123  & 0.3989  & 0.0045  & 0.0027  \\
$n=100$ & max  & 14  & 0.912  & 49  & 99  & 0.3557  & 15.524  & 0.1011  & 0.0365  \\
$m=200$ & mean & 6.47  & 0.389  & 25.3  & 36.1  & 0.1353  & 4.9246  & 0.0562  & 0.0176  \\
\midrule
{4} & min & 6 & 0.718 & 20 & 20 & 0.1110 & 4.2690 & 0.0634 & 0.0164 \\
$n=100$ & max & 36 & 4.959 & 128 & 154 & 0.8094 & 33.2009 & 0.3805 & 0.1197 \\
$m=400$ & mean & 12.6 & 1.488 & 53.6 & 68.6 & 0.3025 & 10.9926 & 0.1601 & 0.0354 \\
\midrule
{5} & min & 4 & 0.825 & 4 & 4 & 0.0701 & 4.9458 & 0.0823 & 0.0140 \\
$n=200$ & max & 17 & 3.876 & 86 & 86 & 1.1996 & 109.7116 & 0.4540 & 0.0794 \\
$m=400$ & mean & 9.4 & 2.164 & 35.8 & 36.6 & 0.4217 & 44.5244 & 0.2223 & 0.0414 \\
\midrule
{6} & min & 2 & 0.669 & 5 & 5 & 0.0625 & 5.4639 & 0.0311 & 0.0052 \\
$n=200$ & max & 24 & 10.279 & 128 & 311 & 4.4925 & 380.9636 & 0.6182 & 0.1250 \\
$m=800$ & mean & 13.4 & 5.379 & 64.6 & 82.4 & 1.1090 & 111.4111 & 0.3150 & 0.0664 \\
\midrule
{7} & min & 4 & 19.586 & 10 & 10 & 3.2243 & -- & 0.4168 & 0.0870 \\
$n=1000$ & max & 18 & 102.092 & 64 & 64 & 18.9144 & -- & 2.3113 & 0.4913 \\
$m=2000$ & mean & 11.8 & 62.721 & 32.6 & 33.4 & 10.8732 & -- & 1.4042 & 0.2709 \\
\midrule
{8} & min & 4 & 39.428 & 7 & 7 & 2.6272 & -- & 0.4336 & 0.0963 \\
$n=1000$ & max & 32 & 341.543 & 112 & 121 & 41.6134 & -- & 4.0621 & 0.9461 \\
$m=4000$ & mean & 18.6 & 195.961 & 55.4 & 61.4 & 22.8661 & -- & 2.5020 & 0.5284 \\
\midrule
{9} & min & 7 & 75.994 & 9 & 9 & 16.1292 & -- & 1.3056 & 0.2225 \\
$n=2000$ & max & 24 & 516.131 & 73 & 124 & 139.9464 & -- & 9.3591 & 1.5732 \\
$m=4000$ & mean & 14.8 & 300.958 & 38.6 & 48.6 & 55.7729 & -- & 5.0278 & 0.8677 \\
\midrule
{10} & min & 9 & 359.654 & 17 & 17 & 27.3013 & -- & 0.5476 & 0.5476 \\
$n=2000$ & max & 26 & 1090.674 & 142 & 220 & 279.7599 & -- & 10.7284 & 1.9579 \\
$m=8000$ & mean & 17.2 & 716.485 & 67.6 & 80.8 & 112.5519 & -- & 6.5519 & 1.1404 \\
\bottomrule
{11} & min & 6 & 777.874 & 14 & 14 & 106.6934 & -- & 11.3873 &  1.4136\\
$n=5000$ & max & 21 & 2741.023 & 110 & 160 & 1010.9983 & -- & 44.9173 & 6.2393 \\
$m=10000$ & mean & 13.6 & 1666.410 & 47.9 & 48.34 & 332.6101 & -- & 26.8930 & 3.6720 \\
\bottomrule
\end{tabular}

 \end{table}

\begin{figure}[htb]
    \centering
    \begin{tikzpicture}[scale=0.8]
    \begin{axis}[
        width=14cm,
        height=8cm,
        ylabel={Computation Time},
        legend pos=north west,
        ymode=log,
        xtick={1,2,3,4,5,6,7,8,9,10},
        xticklabels={1,2,3,4,5,6,7,8,9,10},
        grid=major,
    ]

    \addplot[color=DodgerBlue4,mark=o] coordinates {(1,0.0269) (2,0.0984) (3,0.1353) (4,0.3025) (5,0.4217) (6,1.1090) (7,10.8732) (8,22.8661) (9,55.7729) (10,112.5519) (11,332.6101)};
    \addlegendentry{$T_{AO}$}
     Therefore, we do not include these larger test, due to their practical limitation.
    \addplot[color=DeepPink4,mark=square*]  coordinates {(1,0.5565) (2,1.8382) (3,4.9246) (4,10.9926) (5,44.5244) (6,111.4111)};
    \addlegendentry{$T_{DS}$}
    
    \addplot[color=Chartreuse4 ,mark=triangle*] coordinates {(1,0.0131) (2,0.0399) (3,0.0562) (4,0.1601) (5,0.2223) (6,0.3150) (7,1.4042) (8,2.5020) (9,5.0278) (10,6.5519) (11,26.8930)};
    \addlegendentry{$T_{\varepsilon}$}
    
    \addplot[color=DarkGoldenrod3,mark=diamond*] coordinates {(1,0.0072) (2,0.0181) (3,0.0176) (4,0.0354) (5,0.0414) (6,0.0664) (7,0.2709) (8,0.5284) (9,0.8677) (10,1.1404) (11,3.6720)};
    \addlegendentry{$T_{\text{New-}\varepsilon}$}
    \end{axis}
\end{tikzpicture}
    \caption{Line graph showing the comparison of average computation times for the different methods across the problem instances from~\Cref{tab2}. The x-axis represents different problem instances, while the y-axis (logarithmic scale) indicates the computation time.}
    \label{fig:LineGraphForTable1}
\end{figure}

\begin{table}[htb]\footnotesize

\caption{Numerical results for the different instances of~\Cref{example1}.  Each instance has $|\Y_{EN}|=2$ extreme supported nondominated and  $|\Y_{SN}|=6$ nondominated supported vectors. For the first 5 instances, it holds $M=10$, $L=5$. For the last, it holds $M=1,L=5$. In these instances the adjusted algorithm needs only $|\Y_{SN}|=6$ branches.  }\label{tab3}
\vspace{0.5cm}
\centering
\begin{tabular}{llrrrrrrr}
\toprule
 Class  &     $|\X_{SN}|$ &  T-AO &   T-DS &  T-$\varepsilon$ &  T-New-$\varepsilon$ \\
\midrule
        $n=5$  & 726       & 0.0587  & 0.0012  & 0.0018  & 0.0013  \\
        $n=6$  & 7986      & 0.6490  & 0.0016  & 0.0017  & 0.0012  \\
        $n=7$  & 87846     & 7.8305  & 0.0019  & 0.0223  & 0.0013  \\
        $n=8$  & 966306    & 95.4091 & 0.0025  & 0.0193  & 0.0013  \\
        $n=9$  & 10629366  & 1133.3333 & 0.0028  & 0.0191  & 0.0013  \\
        $n=20$ & 786432    & 165.3812 & 0.0136  & 0.0219  & 0.0020  \\
\bottomrule
\end{tabular}

 \end{table}

\begin{table}[htb]\footnotesize

\caption{Numerical results for the different instances of~\Cref{example2}.  Each instance has $|\Y_{EN}|=2$ extreme supported nondominated and  $|\Y_{SN}|=6$ nondominated supported vectors. For the first six instances, it holds $L=5$. For the last $L=3$. In this instances the adjusted algorithm needs exactly $|\X_{SN}|$ branches.}\label{tab4}
\vspace{0.5cm}
\centering
\begin{tabular}{llrrrrr}
\toprule
 Class  & $|\X_{SN}|$ & T-AO & T-DS & T-$\varepsilon$ & T-New-$\varepsilon$ \\
\midrule
 $n=5$  & 56     & 0.0048  & 0.0120  & 0.0016  & 0.0012  \\
 $n=6$  & 126    & 0.0133  & 0.0356  & 0.0016  & 0.0012  \\
 $n=7$  & 252    & 0.0427  & 0.0864  & 0.0019  & 0.0013  \\
 $n=8$  & 462    & 0.0641  & 0.1954  & 0.0020  & 0.0014  \\
 $n=9$  & 792    & 0.1257  & 0.4160  & 0.0021  & 0.0014  \\
 $n=10$ & 1287   & 0.2221  & 0.8041  & 0.0202  & 0.0016  \\
 $n=20$ & 33649  & 12.9121 & 84.201  & 0.0195  & 0.0027  \\
\bottomrule
\end{tabular}

 \end{table}

 \begin{table}[htb]\footnotesize
 \caption{Displays the time difference of the $\varepsilon$-epsilon method versus the more compact formulation of the mean for each randomly generated NETGEN class instances of~\Cref{tab2}. }\label{tab5}
 \vspace{0.5cm}
\centering
\begin{tabular}{lr}
\toprule
Class & T-$\varepsilon$ / T-New-$\varepsilon$ \\
\midrule
1 & 1.82 \\
2 &  2.20 \\
3 & 3.19 \\
4 & 4.52 \\
5 & 5.37 \\
6  & 4.74 \\
7 & 5.18 \\
8 & 4.74 \\
9 & 5.79 \\
10& 5.73 \\
11& 7.87 \\
\bottomrule
\end{tabular}

\end{table}

\Cref{tab3} indicates, as expected, that the adjusted algorithm outperforms the all optimal flow algorithm when the number of branches needed is significantly smaller than the number of supported efficient solutions. However, when the number of branches needed equals the number of efficient solutions, the all optimal algorithm outperforms the adjusted algorithm, as evident in~\Cref{tab4}. The high computational cost of the Floyd-Warshall algorithm used in each branch contributes to the adjusted algorithm's suboptimal performance, as highlighted in~\Cref{tab2}, and indicates that this algorithm performs poorly in practice. 

Despite this, when the number of supported nondominated vectors and supported efficient solutions equals, and therefore, the all optimal flow algorithm would run in output-polynomial time to determine the number of all supported nondominated vectors, the objective-space methods clearly outperform this algorithm. 

 The observed differences in running times of the outcome space methods against the decision space methods can, in part, be attributed to the nature of the implementations. Gurobi, a highly optimized solver, benefits from extensive engineering and decades of refinement. In contrast, the other algorithms in our study were implemented from scratch in Python, which inherently results in slower execution times due to the interpreted nature of the language and the absence of low-level optimizations.  

It is important to emphasize that developing a high-performance, C-based implementation of our proposed algorithms is beyond the scope of this paper. However, our results still provide valuable insights into the relative efficiency of different approaches independent of implementation-specific optimizations. It is important to note that despite the existence of high-performance implementations for outcome space methods, the development of new decision space methods should not be overlooked.

To foster reproducibility and further research,  our code and benchmark instances can be found in an open repository under the following \href{https://github.com/David-Koenen/All_Supported_Nd_Vectors_BMCIF}{link} to allow other researchers to replicate our findings, test alternative implementations, and explore further optimizations.  

  On a positive note, the more compact formulation of the $\varepsilon$-constraint scalarization surpasses the classic formulation in each instance, as shown in~\Cref{tab5}. The mean CPU time needed in the more compact formulation for the $\varepsilon$-constraint scalarization across all instance classes is  $4.244$ times faster than the standard formulation, demonstrating its efficiency. The time gap may increase in instances including more nodes and arcs. This is a nice result since the more compact formulation can also be used in any $\varepsilon$-constraint scalarization method for multi-objective minimum cost integer flow problems with $d \geq 3$ objectives or the determination of all nondominated vectors for minimum cost integer flow problems.
 
 In conclusion, the computational experiments provide valuable insights into the strengths and weaknesses of the implemented algorithms. The more compact formulation of $\varepsilon$-constraint scalarization presents a promising direction for future research, offering faster computation times and potential applications in multi-objective minimum cost integer flow problems with three or more objectives.

\section{Conclusion}\label{chapt:concl}
This paper discusses the time complexity of enumerating all supported nondominated vectors for MMCIF. The paper shows that there cannot exist an output-polynomial time algorithm for the enumeration of all supported nondominated vectors that determine the vectors in an ordered manner in the outcome space unless $\mathbf{P}=\mathbf{NP}$. However, the question of whether an output-polynomial time algorithm exists remains open. Future research could focus on whether an output-polynomial time algorithm exists for the bi-objective or the multi-objective case with $d\geq 3$ objectives. 
The numerical tests show that the outcome space methods clearly outperform the decision-space methods, even if they do not run in output-polynomial time. The compact formulation of the ILP for the $\varepsilon$-method shows a significant time improvement compared to the conventional ILP. The new formulation can also be used to compute all nondominated vectors in multidimensional minimum cost integer flow problems and could be investigated in the future. 
    
For MMCIF, supported nondominated vectors are often only a minor part of the complete set of nondominated vectors, even for the bi-objective case. However, determining supported nondominated vectors is often needed as a first step in two-phase exact methods and for population-based heuristics.  The development of improved two-phase methods that compute all nondominated vectors for MMCIF could be explored in the future.

\bigskip

\textbf{Acknowledgements.} David Könen acknowledges financial support from Deutsche Forschungsgemeinschaft within the
project number 441310140.

\end{document}